\documentclass{aa}  

\usepackage{graphicx}
\usepackage{txfonts}
\usepackage[colorlinks=true,allcolors=magenta]{hyperref}
\usepackage{amsmath}
\usepackage{calrsfs}
\usepackage{physics}
\DeclareMathAlphabet{\pazocal}{OMS}{zplm}{m}{n}
\newcommand{\XMM}{\textit{XMM-Newton}}
\newcommand{\Chandra}{\textit{Chandra}}
\newcommand{\La}{\mathcal{L}}
\newcommand{\LX}{L_\mathrm{X}}
\newcommand{\mnh}{N_\mathrm{H}}
\newcommand{\nh}{$N_\mathrm{H}$}
\newcommand{\dlog}{\mathrm{dlog}}
\newcommand{\pden}{{p_\mathrm{den}}}
\newcommand{\fabs}{f_\mathrm{abs}}
\newcommand{\fctk}{f_{\mathrm{CTK},r}}
\newcommand{\Lbol}{L_\mathrm{bol}}

\begin{document} 
   \title{AGN X-ray luminosity function and absorption function in the Early Universe ($3\leq z \leq 6$)}

   \subtitle{}
    \titlerunning{AGN X-ray luminosity function and absorption function in the Early Universe }
    \authorrunning{E. Pouliasis et al.}

    \author{E.~Pouliasis\inst{1}\thanks{E-mail: epouliasis@noa.gr} 
    \and A.~Ruiz\inst{1}
    \and I.~Georgantopoulos\inst{1}
    \and F.~Vito\inst{2}
    \and R.~Gilli\inst{2}
    \and C.~Vignali\inst{3,2}    
    \and Y.~Ueda\inst{4}
    \and E.~Koulouridis\inst{1}
    \and M.~Akiyama\inst{5}
    \and S.~Marchesi\inst{3,2,6}
    \and B.~Laloux\inst{1,7}
    \and T.~Nagao\inst{8}
    \and S.~Paltani\inst{9}     
    \and M.~Pierre\inst{10}
    \and Y.~Toba\inst{11,12,8}
    \and M.~Habouzit\inst{13,14}
    \and B.~Vijarnwannaluk\inst{5,12}
    \and C.~Garrel\inst{10}
}

    \institute{IAASARS, National Observatory of Athens, Ioannou Metaxa and Vasileos Pavlou GR-15236, Athens, Greece\\
    \email{epouliasis@noa.gr}
    \and
    INAF - Osservatorio di Astrofisica e Scienza dello Spazio di Bologna, Via Gobetti 93/3, I-40129 Bologna, Italy
    \and
     Università di Bologna, Dip. di Fisica e Astronomia ``A. Righi'', Via P. Gobetti 93/2, I-40129 Bologna, Italy
    \and
    Department of Astronomy, Kyoto University, Kitashirakawa-Oiwake-cho, Sakyo-ku, Kyoto 606-8502, Japan
     \and
    Astronomical Institute, Tohoku University, 6-3 Aramaki, Aoba-ku, Senda, 980-8578, Japan
    \and
    Department of Physics and Astronomy, Clemson University,  Kinard Lab of Physics, Clemson, SC 29634, USA
    \and
    Centre for Extragalactic Astronomy, Department of Physics, Durham University, UK
    \and
    Research Center for Space and Cosmic Evolution, Ehime University, 2-5 Bunkyo-cho, Matsuyama, Ehime 790-8577, Japan
    \and
    Department of Astronomy, University of Geneva, ch. d'Écogia 16, CH-1290 Versoix, Switzerland
    \and
    AIM, CEA, CNRS, Université Paris-Saclay, Université Paris Diderot, Sorbonne Paris Cité, F-91191 Gif-sur-Yvette, France
    \and
    National Astronomical Observatory of Japan, 2-21-1 Osawa, Mitaka, Tokyo 181-8588, Japan
    \and
    Academia Sinica Institute of Astronomy and Astrophysics, 11F of Astronomy-Mathematics Building, AS/NTU, No.1, Section 4, Roosevelt Road, Taipei 10617, Taiwan
    \and
    Zentrum f\"{u}r Astronomie der Universit \"{a}t Heidelberg, ITA; Albert-Ueberle-Str. 2, D-69120 Heidelberg, Germany 
    \and
    Max-Planck-Institut f\"{u}r Astronomie, K\"{o}nigstuhl 17, D-69117 Heidelberg, Germany
    }

   \date{Received  ; accepted }

% \abstract{}{}{}{}{} 
% 5 {} token are mandatory

  \abstract
   {The X-ray luminosity function (XLF) of active galactic nuclei (AGN) offers a robust tool to study the evolution and the growth of the super-massive black-hole population over cosmic time. Owing to the limited area probed by X-ray surveys, optical surveys are routinely used to probe the accretion in the high redshift Universe $z\geq 3$. However, optical surveys may be incomplete because they are strongly affected by dust redenning. 
   In this work, we derive the XLF and its evolution at high redshifts ($z\geq 3$) using a large sample of AGNs selected in different fields with various areas and depths covering a wide range of luminosities. Additionally, we put the tightest yet constraints on the absorption function in this redshift regime.
   In particular, we use more than 600 soft X-ray selected ($0.5-2$ keV) high-z sources in the Chandra Deep fields, the Chandra COSMOS Legacy survey and the XMM-XXL northern field. We derive the X-ray spectral properties for all sources via spectral fitting, using a consistent technique and model. For modeling the parametric form of the XLF and the absorption function, we use a Bayesian methodology allowing us to correctly propagate the uncertainties for the observed X-ray properties of our sources and also the absorption effects.
   The evolution of XLF is in agreement with a pure density evolution model similar to what is witnessed at optical wavelengths, although a luminosity dependent density evolution model cannot be securely ruled out. A large fraction ($\sim 60 \%)$ of our sources are absorbed by column densities of $\rm N_H \geq 10^{23} cm^{-2} $, while $\sim 17$\% of the sources are Compton-thick. Our results favor a scenario where both the interstellar medium of the host and the AGN torus contribute to the obscuration. The derived black hole accretion rate density is roughly in agreement with the large-scale cosmological hydro-dynamical simulations, if one takes into account the results that the X-ray AGN are hosted by massive galaxies, while it differs from the one derived using JWST data. The latter could be due to the differences in the AGN and host-galaxy properties.}

%  while it lies below the star-formation density suggesting that an additional fraction of AGN may be associated with Compton-thick AGN.

   \maketitle
%
%-------------------------------------------------------------------
\section{Introduction}
\label{intro}
In the last several years it has become clear that most galaxies have a super-massive black hole (SMBH) in their centre \citep[e.g.][]{KormendyHo2013}. SMBHs are active, dubbed as active galactic nuclei (AGN), if there is material falling towards the centre of their galaxies. In most cases, the in-falling material creates a geometrically thin optically thick \citep{ShakuraSunyaev1973} accretion disk producing copious amounts of radiation in the extreme UV part of the spectrum. Alternatively, the accretion may be radiatively inefficient channeling large amounts to the production of jets. All AGN produce intense X-ray radiation which equals roughly a few percent of the bolometric luminosity \citep{Lusso2012,Duras2020}. The X-ray emission is believed to originate from Compton up-scattering of the ultraviolet (UV) accretion disk photons \citep[e.g][]{HaardtMaraschi1991} on a hot electron corona with a mean temperature of $\rm k$$T_\mathrm{e}\sim 65\pm10\,\mathrm{keV}$ \citep{AkylasGeorgantopoulos2021,Kamraj2022}. 

It is evident that the ubiquitous X-ray emission provides one of the most robust ways to detect AGN \citep{BrandtAlexander2015}. This is because X-ray wavelengths are hardly affected by obscuration \citep{Burlon2011,HickoxAlexander2018,Georgantopoulos2019,Mountrichas2020,Georgakakis2020,Toba2022}. Moreover, in most cases the contaminating star formation contributes a small fraction of the X-ray emission. The deepest X-ray observations in the \Chandra\ deep field \citep{Luo2017} have revealed a surface density of about 30\,000 sources per square degree where AGN form the vast majority of these sources. In contrast, the optical AGN surveys, e.g., the albeit much shallower ($g<22.5$ mag) Sloan Digital Sky Survey \citep[SDSS,][]{Paris2018}, reach a surface density of less than 200 luminous broad line AGN per square degree. 

The \Chandra\ \citep{Weisskopf2000} and \XMM\ \citep{Jansen2001} X-ray surveys, in the $2-10$~keV band, have allowed us to study in detail the AGN demographics up to redshifts of $z\sim3-4$ \citep[e.g.,][]{Ueda2003,Ueda2014,Aird2015, Ranalli2015, Miyaji2015,Buchner2015,Georgakakis2017,Peca2023,Laloux2023}. The AGN luminosity function can be well described by a double power-law which evolves with redshift according to a luminosity dependent model \citep[e.g.][]{Ueda2014}. According to this model, the AGN evolution follows a 'cosmic downsizing' pattern in the sense that the most luminous AGN ($\rm \log L_X(2-10keV)[erg~s^{-1}]= 45-47$) have been formed first ($z\sim2$), while the less luminous ones  $\rm (\log L_X(2-10keV)[erg~s^{-1}]= 42-43$) have the peak of their redshift distribution at redshifts below one \cite[e.g,][]{Ueda2014,Aird2015}. The above authors compare the evolution of the galaxy star-formation rate density and the black hole accretion density (BHAD) as derived from the X-ray luminosity function. Although both peak at $z\sim 2$, at higher redshifts ($z=4-5$), the BHAD presents a much stronger decline over redshift compared to the star formation rate density (SFRD). Thus, galaxy growth may precede the build up of their central SMBHs in the early Universe \citep{Aird2015}. Alternatively, the SMBH may have formed massive enough and thus they do not need high accretion rates to reach the local $\rm M_{BH}-M\star$ relation. However, the X-ray data are quite scarce at these redshifts and therefore this result awaits confirmation.

At redshifts higher than $z=3$, X-ray surveys have harvested limited AGN samples because of the limited sky area covered. The density of high-redshift AGN is very low \citep[$\rm \sim 1~Gpc^{-3}$,][]{DeRosa2014} and therefore large areas need to be probed. The X-ray luminosity function at high redshifts ($3<z<6$) has been derived by \cite{Vito2014}, \cite{Vito2018}, \cite{Georgakakis2015}. \cite{Vito2018} using about one hundred X-ray sources from the CDF-N and CDF-S, focused on the faint end of the luminosity function, finding a very high fraction of obscured sources $\sim0.6-0.8$ with column densities $\rm \log N_H [cm^{-2}]\geq 23$. The ongoing all-sky extended ROentgen Survey with an Imaging Telescope Array \citep[eROSITA,][]{Predehl2021} is expected to facilitate the search for high redshift AGN owing to its large grasp (field-of-view multiplied by effective area). After four years of operation, a few examples of very luminous $z>6$ AGN have been identified with the eROSITA detector \citep{Wolf2021,Wolf2023,Medvedev2020} with the majority of them being radio-loud. The X-ray telescope onboard the {\it Gehrels/SWIFT} mission  has provided yet another $z>6$ AGN \citep{BarlowHall2023}. The serendipitous \XMM\ catalogue \citep{Webb2020} provides another rich resource for detecting high-redshift AGN. Until December 2022, 657,000 unique sources had been detected covering an area of about $1300\deg^2$. The next release of the \XMM\ serendipitous source catalogue \citep{Webb2023} is expected to provide photometric redshifts derived using deep optical photometry, e.g. from the Dark Energy Survey \citep{Abbott2021}. In the near future, deep near-IR data from the {\it EUCLID} mission \citep{Scaramella2022} will help to provide even more accurate photometric redshifts at $z>4$. Nevertheless, the redshift confirmation of X-ray selected sources still requires the spectroscopic follow-up of the optical counterpart. This task is particularly difficult at high redshift, because of the faintness of the optical counterparts.

In contrast to the relatively limited advances at X-ray wavelengths, at redshifts $z>3$, the optical surveys have discovered high numbers of broad line AGN. This is because of the availability of wide-field (i.e., $\sim10^4\deg^2$) optical/near-infrared (NIR) surveys, such as the SDSS \citep{Jiang2016}, the UKIRT Infrared Deep Sky Survey (UKIDSS; \cite{Mortlock2011}), the Canada-France High-redshift Quasar Survey (CFHQS;  \cite{Willott2010}) and the Panoramic Survey Telescope \& Rapid Response System \citep{Banados2018}. These led to the discovery of more than 300 broad line AGN at $z > 5.8$ \citep{Fan2023} when the Universe was less than one Gyr old. The highest redshift AGN discovered by the above surveys was identified at $z=7.642$ \citep{Wang2021}. Deep optical surveys with the Subaru Hyper Supreme Cam, HSC, \citep{Miyazaki2017} such as the HSC Subaru Strategic Plan Survey \citep{Aihara2018,Aihara2019} allowed the detection of AGN at redshifts $z=3-6$ at much fainter ($\rm >3~mag$) absolute magnitudes \citep{Akiyama2018,Niida2020,Matsuoka2022}. The optical luminosity function decreases rapidly above redshifts $z\sim3$. The drop in the AGN density is consistent with a pure density evolution model \citep{Mcgreer2013, Matsuoka2023}. Recently, the launch of the {\it James Webb Space Telescope} (\textit{JWST}) allowed the detection of faint AGN up to redshifts of $\sim 10$ \citep[e.g.][]{Kocevski2023,Yang2023,Castellano2023, Bogdan2023}. However, it is likely that the optical/UV luminosity function may be affected by large amounts of dust attenuation. When \citet{Lusso2023} convert the UV luminosity function to the X-ray band, they find that the X-ray luminosity function in the redshift range $z=3-6$ is almost an order of magnitude higher than the optical. In this high-redshift regime, it is likely that both the obscuring torus and the interstellar medium contribute to the obscuration \citet{Gilli2022}. Interestingly, recent {\it JWST} mid-IR observations indicate that even the X-ray surveys may be affected by Compton-thick obscuration. \cite{Yang2023}, using the Mid-Infrared Instrument (MIRI) onboard {\it JWST}, suggest a black hole accretion density (BHAD) which is $\sim 0.5$ dex higher than the X-ray results at $z=3-5$.  At even higher redshifts, the obscuration should be even higher. The {\sc BLUETIDES} large volume cosmological simulations \citep{Ni2020} show that at $z>7$ a large fraction of AGN (0.6-1) could be heavily obscured by column densities $N_{\rm H} \rm \geq 10^{23} cm^{-2}$.

Here, we visit anew the X-ray luminosity function at high redshifts ($z=3-6$). We combine the most sensitive observations in X-rays in the \Chandra\ deep fields with the COSMOS-Legacy $2\deg^2$ \Chandra\ observations and the large area of the \XMM/XXL survey ($25\deg^2$). Our sample is the largest ever assembled in X-rays. It contains over 600 sources, including 100 and 30 sources above redshift z=4 and z=5, respectively. The selection of the high-z sample is presented in Sect.~\ref{data}. In Sect.~\ref{xrayproperties}, we analyze the X-ray properties of our sample, and in Sect.~\ref{xlaf} we explain the methodology and the models we used to constrain the X-ray luminosity function and the absorption function. Section~\ref{results} compares our results with other X-ray studies in the literature and also the predicted values coming from the theoretical simulations. Then, we constrain the space density and the black hole accretion density. In Sect.~\ref{conclusions} we summarise the results. Throughout the paper, we assume a standard $\Lambda$CDM cosmology with $\mathrm{H_0=70\, km\, s^{-1}\,Mpc^{-1}}$, $\Omega_\mathrm{M}=0.3$ and $\Omega_\Lambda=0.7$ \citep{Komatsu2009}.

\section{Sample selection}\label{data}

\begin{figure}
\center
    \includegraphics[width=0.48\textwidth]{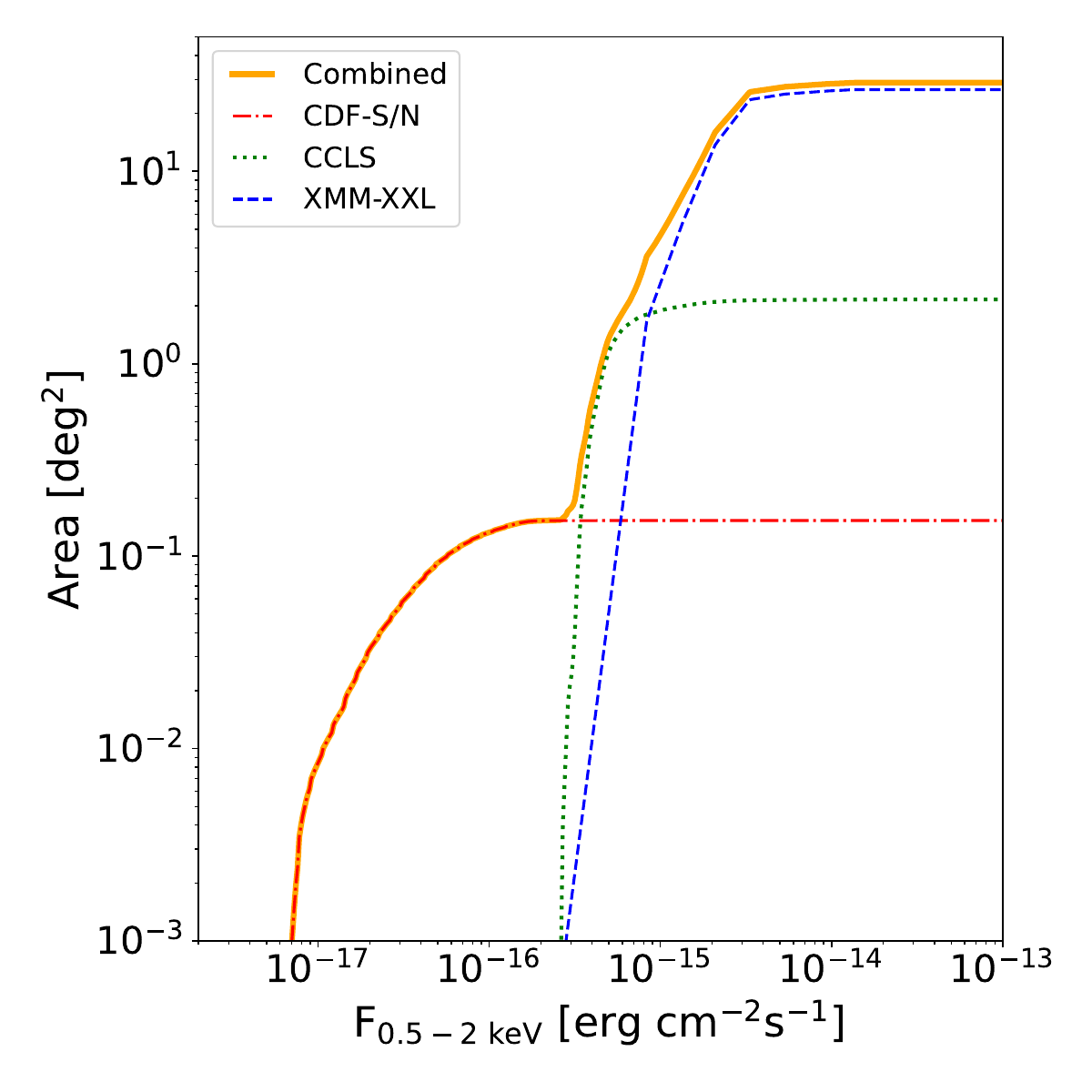} 
\caption{X-ray sensitivity curves presented individually for the CDF-S/N, CCLS and XMM-XXL-N fields. The total area curve is shown with the orange solid line.}
\label{area_curves}
\end{figure}

We derive the X-ray luminosity function in the rest-frame $\rm 2-10~keV$ band and the absorption function of $z\geqslant3.0$ AGN. Since the observed $\rm 0.5-2~keV$ band corresponds to a rest-frame $\rm 2-8~keV$ band at a redshift $z=3$ and to the $\rm 3-12~keV$ band at a redshift of $z=5$, we construct our high-z sample using the soft-band detected sources. We select our sample using three different X-ray surveys: the \Chandra\ Deep Fields: South \citep{Luo2017} and North \citep{Xue2016}, the \Chandra\ COSMOS Legacy Survey \citep[CCLS,][]{Civano2016} and the northern region of the \XMM\ XXL survey \citep[][XMM-XXL-N]{pierreXXL}. These surveys cover various sky areas and depths, allowing for the compilation of a high-z data set with the highest possible completeness with respect to luminosity, redshift and absorption column density ranges. The sensitivity-area curves for these surveys are shown in Fig.~\ref{area_curves}. The three surveys probe a large range of fluxes, allowing us to cover luminosities which span four orders of magnitude. Below, we give a brief description of the high-redshift AGN selection used in each field.

\subsection{X-ray selected AGN from the XMM-XXL northern field}

XMM-XXL-N covers an area of about $25\deg^2$ at a depth of $\rm 6 \times10^{-15}  \,erg~cm^{-2}~s^{-1}$ (at 3$\rm \sigma$) in the soft band ($\rm 0.5-2 keV)$. Parts of this area have been observed in the framework of the XMM-SERVS survey \citep[$5.4\deg^2$,][]{Chen2018} and the Subaru/XMM-Newton Deep Survey \citep[$1.3\deg^2$,][]{Ueda2008}, with sensitivity limits in the soft band of $\rm 1.7 \times 10^{-15}$ and $\rm 6 \times 10^{-16} ~ergs~cm^{-2}~s^{-1}$, respectively. We used the internal release of the XMM-XXL catalogue obtained with the V4.2 XXL pipeline that contains in total 15547 X-ray sources. Restricting our area to the Hyper Suprime-Cam Subaru Strategic Program \citep[HSC-SSP,][]{Miyazaki2018} coverage, we ended up with 10232 soft-band detected sources. This reduced the area to $\sim 21~deg^2$. In \citet{Pouliasis2022a}, we presented a catalogue of high-z AGN using the HSC colour-colour diagrams which are based on the Lyman break (drop-out) techniques. However, the above sample was limited to sources with $z\geq3.5$ due to the lack of coverage in the u band. Furthermore, the Lyman-break method may miss a fraction of the high-redshift sources due to the different morphological selections or because of the redenned or host-galaxy dominated colours in the case of the obscured AGN \citep{LeFevre2005,Paltani2007,Boutsia2021,klod2022}. Thus, in order to increase the completeness of our sample we run the LePHARE algorithm \citep{Arnouts1999, Ilbert2006} for all the soft-band X-ray sources in XMM-XXL-N. In Appendix \ref{lephare}, we provide the whole procedure followed to estimate the photometric redshifts and gather all the available spectroscopic information. Concerning the photo-z sample, we found that the fraction of outliers is $\eta$=20.9\% and the scatter between spectroscopic and photometric redshift is $\sigma_{\rm NMAD}$=0.07. These statistics are similar or slightly better compared to previous X-ray studies \citep[e.g,][]{Salvato2022,klod2022}. The final XMM-XXL-N sample of high-z sources includes 70 sources with spectroscopic redshifts and 438 sources that have a probability of more than 20\% to be at $z\geq3$ in their PDF(z). Among them, 319 sources have photometric redshifts with $z\geq3$. Taking into account the sum of the PDF(z) of all sources at $z\geq3$ in addition to the sources with spectroscopic redshift, the effective number of high-z sources is 390.7.

\subsection{X-ray selected AGN from the CCLS}

CCLS covers an area of $2.2\deg^2$ with a mosaic of $\sim180$~ks \Chandra\ pointings, for a total observing time of about 4.6\,Ms, reaching a depth of $\rm 2.2\times 10^{-16} erg~cm^{-2}~s^{-1}$ in the soft X-ray band (0.5-2 keV). \citet{Marchesi2016optical} provided optical and infrared identifications for the whole sample of 4016 X-ray sources in the CCLS and obtained photometric redshifts using the LePHARE SED fitting code. \citet{Marchesi2016highz} built a sample of 174 sources within redshift $\rm 3<z<6.85$. 147 of them are detected in the soft band and were used in our work. Among them, 81 have available spectroscopic redshift. Additionally, we included lower-redshift sources which have a probability of more than 20\% to lie at high redshift. The effective number of soft-band detected high-z sources is 151.2 in the CCLS field.

\subsection{X-ray selected AGN from the \Chandra\ deep fields}

The \Chandra\ deep fields (CDF), namely CDF-South and CDF-North, cover a total area of $0.15\deg^2$. These are the deepest observations available in X-rays so far, reaching a depth of $\rm 6.4\times 10^{-18}~erg~cm^{-2}~s^{-1}$ and $\rm 1.2\times 10^{-17}~erg~cm^{-2}~s^{-1}$ in the soft band, respectively. In this work, we made use of the high-redshift AGN catalogue produced by \citet{Vito2018}. They have gathered all the information for redshift estimates and provided a catalogue with X-ray sources having a probability larger than 20\% of being in the high redshift regime, $z\geq3$. They only focused on the central regions of both fields, where vignetting effects and distortions of the point spread function (PSF) are minimal, and hence reaching an effective exposure above 1~Ms. The final area used was reduced ($\rm \sim330~arcmin^2$ and $\rm \sim215~arcmin^2$ in CDF-S and CDF-N, respectively). Since newer spectroscopic surveys and photometric data became available, we searched whether there are new derived spectra for these sources. In the south, we cross-matched (using the optical counterparts of the X-ray sources and a search radius of 1") the X-ray catalogue with the final data release (DR4) of the VANDELS deep public ESO spectroscopic survey \citep{Pentericci2018,Mclure2018,Garilli2021} and also with the MUSE (Multi Unit Spectroscopic Explorer) Hubble Ultra Deep Field (HUDF) survey \citep{Inami2017}. Regarding CDF-North, we examined the sources with several catalogues in the literature (e.g., \citet{Momcheva2016}). At the end, using the new spectroscopic information, we updated the previous photometric redshifts for twelve high-z sources. Ten out of twelve sources remained in the high-z regime, while we excluded two sources for which the new spec-z was below $z<3$. In addition, we included in the sample six new high-z sources. In total, our sample contains 93 sources with $z\geq3$. Among them, 40 sources have spectroscopic redshifts and 53 sources have photometric redshift estimations. Taking into account the full PDF(z) of the sources that lack spectroscopic information, in addition to the spec-z sample, the effective number of high-z AGN is 89.3.

\subsection{Summary of the high-z sample}
\begin{figure}
     \includegraphics[width=0.47\textwidth]{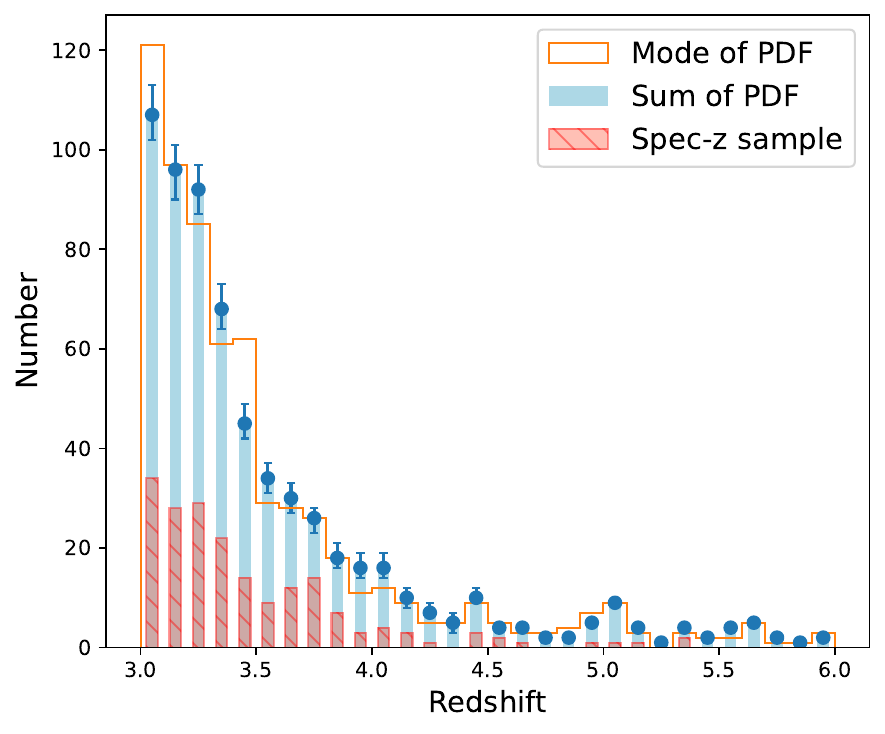}
\caption{Redshift distribution of our sample. The blue bars correspond to the sum of the PDF(z). The PDF(z) of sources with available spectroscopic redshift are represented by a Delta function centered at the spec-z value. The orange line represent the redshift histogram when taking into account only the mode of the PDF(z) for each source. The red hatched bars represent the spec-z sample.} 
\label{redshift}
\end{figure}

\begin{figure}
    \includegraphics[width=0.48\textwidth]{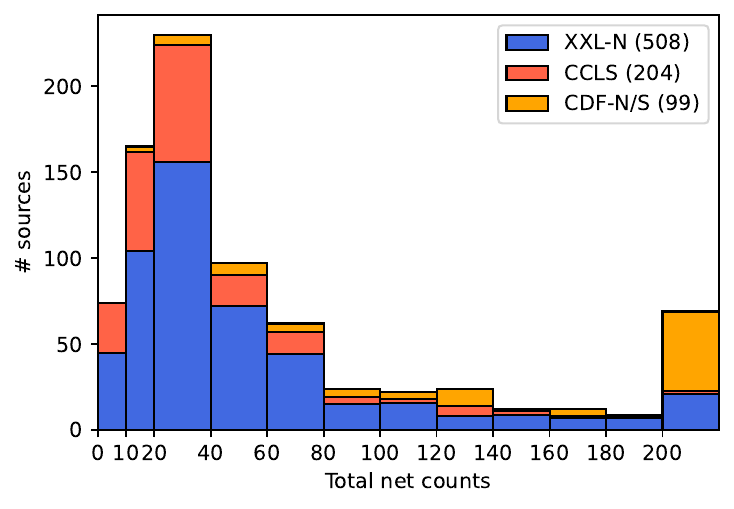}
\caption{Distribution of net counts (i.e. total counts minus background counts) for all the sources included in our spectral analysis. For \XMM~spectra the counts correspond to the 0.3-10 keV interval, while for \Chandra~spectra correspond to 0.3-8 keV. The last bin contains all sources with more than 200 counts.} 
\label{netcountsdist}
\end{figure}
Table~\ref{numbercounts} summarises the final numbers of high-z AGN in different redshift bins selected through our analysis. We also report the numbers derived in each field individually. There are in total 191 sources with secure spectroscopic redshift at $z\geqslant3$, 19 sources at $z\geqslant4$ and four sources with redshift $z\geqslant5$. Using the derived probability density function of the photometric redshifts, PDF($z$), we were able to select additionally 438 sources where the maximum of the PDF($z$) is above 3. However, a closer look at the PDF($z$) of these sources reveals that, while for most of them the probability is concentrated around a narrow redshift range, a non-negligible number of sources show broad or doubled-peaked distributions. Hence, basing our selection only on the maximum probability value would be too restrictive, leaving out of our sample a number of sources with a high likelihood of being at high redshift. We therefore decided to include all sources for which $\mathrm{P}(z\geqslant3) > 0.2$, i.e., there is at least a 20 percent chance for the source to be at high redshift. This threshold has been adopted by \citet{Vito2018} to prevent including sources that show long tails with extremely low probability. A higher-cutoff, such as 0.5, excludes about 10\% of the effective number of sources that are equally distributed in the redshift range of our analysis. Even in this case, the differences in the XLF and the absorption function results are negligible. Furthermore, since luminosity priors have been applied in the photometric redshift estimation procedure in all fields, we are confident that the probability threshold used does not overestimate the number counts in the high-z regime. Finally, taking into account all the above, the effective number count is $\sim631.2$. Our sample of X-ray selected sources is the largest to date in the early Universe ($3\leq z\leq 6$).

\begin{table*}
\caption{Number of sources in different redshift bins for each field and ensemble. We report the number of sources with spectroscopic redshifts and also the total number of high-z sources that is the spec-z sample in addition to the sum of PDF(z) of the photometric redshift sample in the redshift range $3.0 \leq z \leq 6.0$.}      
\label{numbercounts}     
\centering                                      
\begin{tabular}{c  | c c |  c c | c c | c c}          
\hline\hline                        

&\multicolumn{2}{c|}{$3\leq z\leq 4$}&\multicolumn{2}{c|}{$4\leq z\leq 5$}&\multicolumn{2}{c|}{$5\leq z\leq 6$}&\multicolumn{2}{c}{$3\leq z\leq 6$} \\

Field  & z$\mathrm{_{spec}}$  & Total & z$\mathrm{_{spec}}$    & Total & z$\mathrm{_{spec}}$   & Total & z$\mathrm{_{spec}}$   & Total \\

\hline

\rule{0pt}{2ex}    
% inserts single horizontal line
    CDF-S/N     & 38 & 75.1  &      1 & 11.6  &        1 &2.6  &        40 & 89.3      \\
    CCLS        & 70 & 128.5  &      9 &   18.3&       2 &   4.4&        81 &   151.2      \\
    XMM-XXL-N   &  64 & 326.3 &      5 &   40.3&      1 &   24.1&        70 &   390.7     \\ 
    \hline                        
\rule{0pt}{2ex}    
    Combined    &  172 & 529.9  &    15 &  70.2 &    4 & 31.1  &        191 & 631.2     \\ 
\hline                                             %inserts single line
\end{tabular}

\end{table*}

Figure~\ref{redshift} presents the redshift distribution of the sample used in our analysis that is the sum of the PDF(z) of each source. The PDF(z) of sources with available spectroscopic redshift are represented by a Delta function centered at the spec-z value. The uncertainties correspond to 1-$\sigma$ confidence level. For reference, we over-plot the redshift histogram only considering the peaks of the PDF($z$) in addition to the spec-z sample. The agreement between the two distributions is very good at all redshifts indicating that the majority of the photometric redshifts show narrow peaks in their PDF($z$). The red-hatched histogram shows the distribution of the spec-z sample.

%%%%%%%%%%%%%%%%%%%%%%%%%%%%%%%%%%%%%%%%%%%%%%%%%%%%%%%%%%%%%%%%%%%%%%%%%%%%%%%%%%%%%%%%%

%--------------------------------------------------------------------

%-----------------------------------------------------------------

\section{X-ray properties of the high-z AGN sample}\label{xrayproperties}

\begin{figure}
\centering
\includegraphics[width=\hsize]{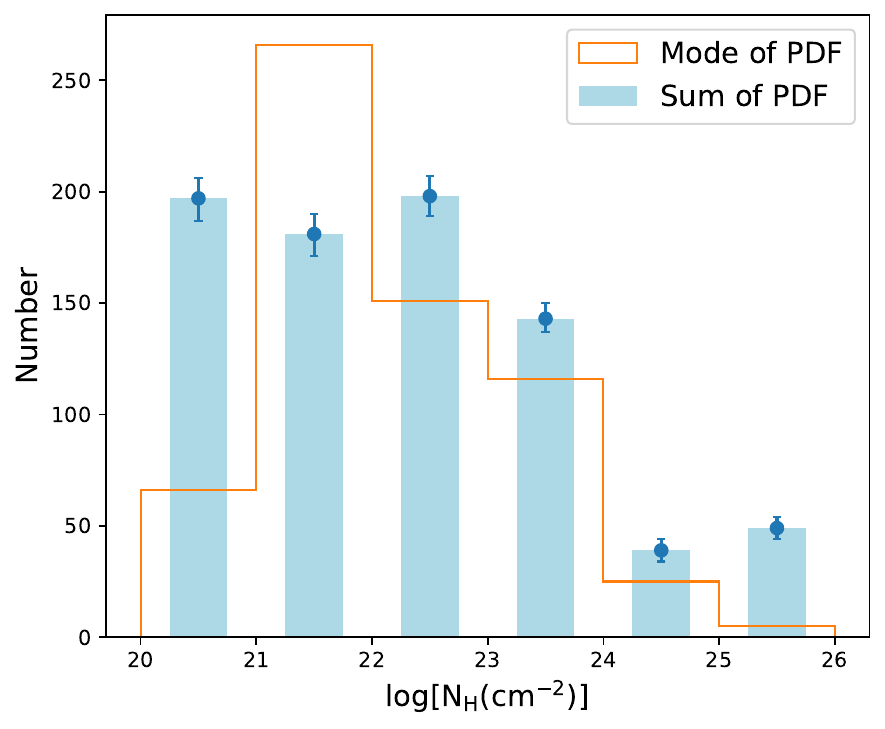}\\
\includegraphics[width=\hsize]{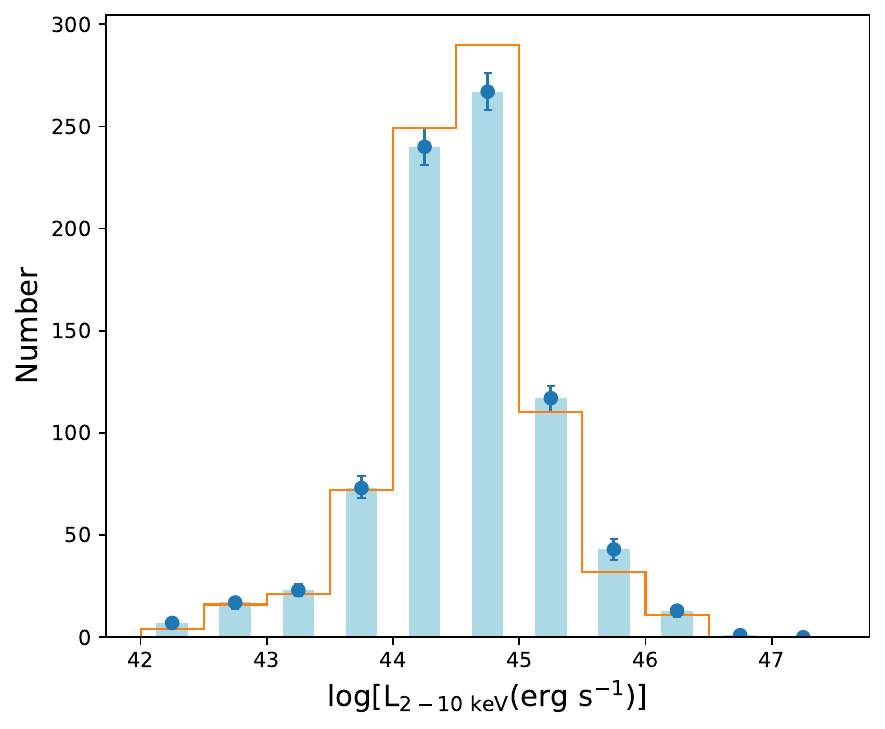}
\caption{Distributions of the Hydrogen column density (top panel) and the $2-10$~keV absorption corrected luminosity (bottom panel) for the high-z sample. The blue bars correspond to the sum of the probability density functions, while the orange lines represent the histograms of the properties when taking into account only the nominal values (mode of the posterior probability distributions).}
\label{distributions}
\end{figure}

\begin{figure*}
\centering
\begin{tabular}{cc} 
    \includegraphics[width=0.47\hsize]{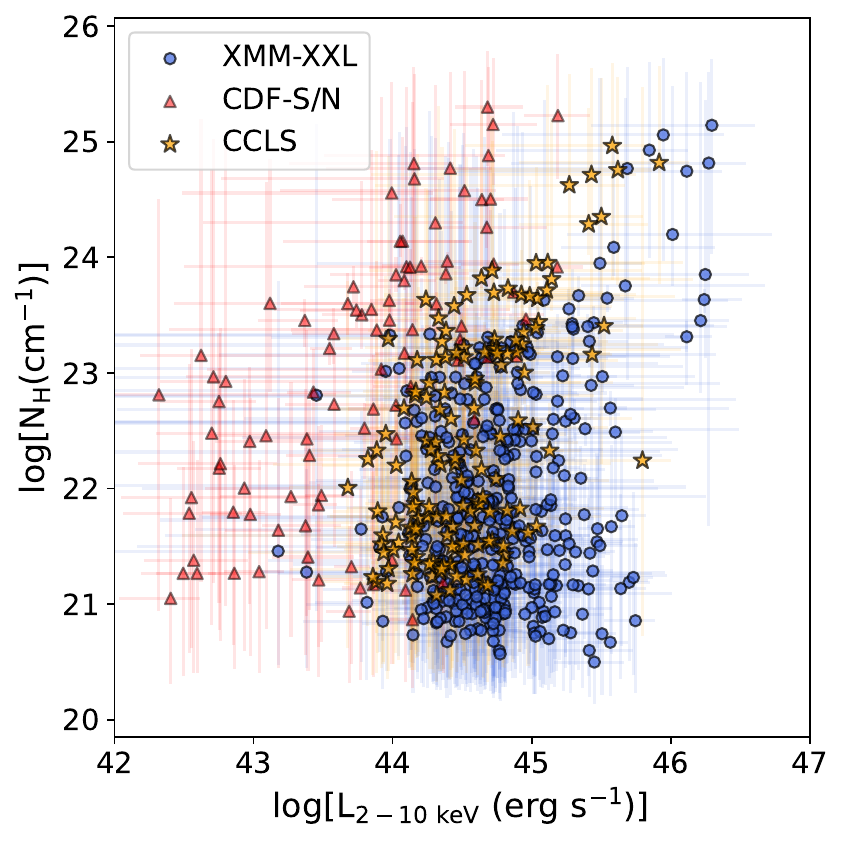} 
&
    \includegraphics[width=0.47\hsize]{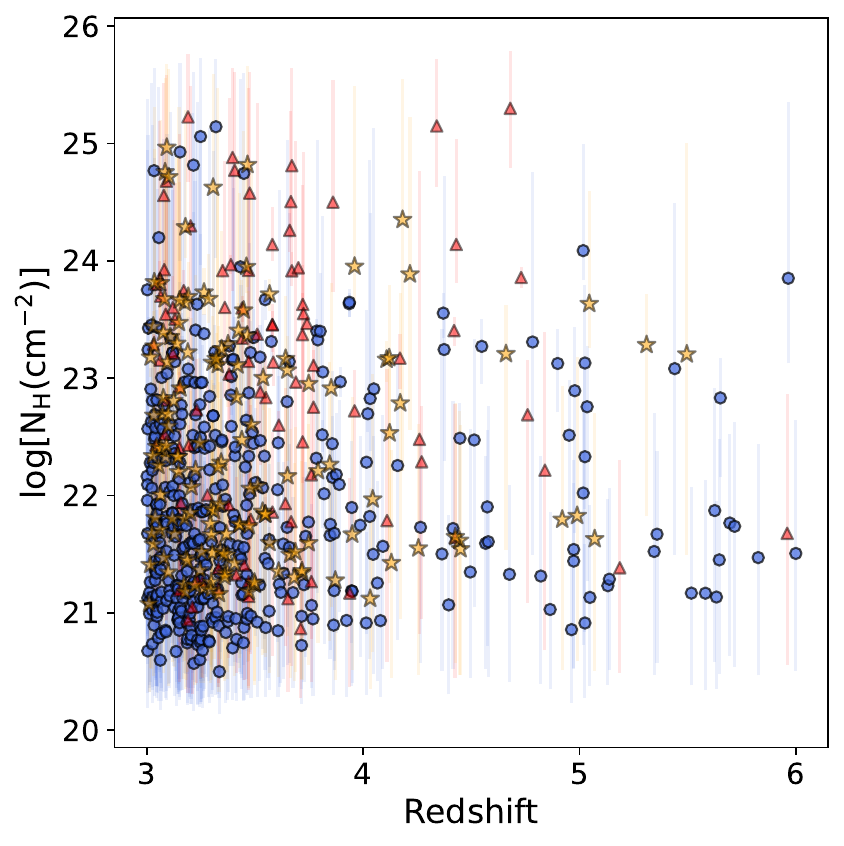} 
\end{tabular}
\caption{Hydrogen column density versus X-ray absorption-corrected, rest-frame luminosity (left) and redshift (right) for the X-ray sources detected in the various fields as indicated in the legend.}
\label{z-lx}
\end{figure*}

In order to calculate the X-ray luminosity and absorption functions, we need accurate estimations of the X-ray properties of the sources in our sample. In particular, we need to estimate the Hydrogen column density, \nh, and the intrinsic, absorption-corrected luminosity in the rest-frame $2-10$\,keV band, $L_\mathrm{X}$. Previous studies of the surveys used for building our high-z sample presented their own X-ray analysis for most of these sources \citep[e.g.,][]{Vito2018,Marchesi2016,Marchesi2016optical,Liu2017}. However, each study used their own methodology (different models and fitting methods) and, in some cases, the X-ray properties for low-count sources are not derived via spectral analysis, but using hardness-ratios (HR). The latter, although useful for identifying highly absorbed sources, introduces high uncertainties, particularly in the \nh\ estimation and hides the underlying assumptions to convert between HR and \nh. However, the use of Cash statistic, which takes into account the Poisson nature of the data, in combination with modern, advanced fitting methods, such as Bayesian X-ray Analysis \citep[BXA,][]{Buchner2014}, allows to handle low-count spectra accurately. Therefore, we have extracted X-ray spectra for all the sources in our sample and analyzed them using an homogeneous approach.

Figure~\ref{netcountsdist} shows the distribution of total net counts for the X-ray spectral products we obtained following the procedures described in Sect.~\ref{xspectra}. The distribution shows the total number of counts, background subtracted, for each source included in our final sample. In the case of sources with multiple spectra from different observations and/or cameras, all counts are added. The plot shows a significant fraction of sources with less than 20 net counts. As stated above, the BXA-based methodology we outline in Sect.~\ref{xfit} allows for a rigorous statistical treatment even in the case of low-count sources, while extracting the maximum amount of information from the available observational data. For such spectra, most of the model parameters are not well constrained and their posterior distributions follow the selected initial priors. Since we use broad, non-informative priors, we do not impose any restrictive values and they are treated as the remaining sources. The large uncertainties of the spectral properties of these sources are then propagated to derived quantities like the X-ray luminosity, and fully taken into account in our methodology for estimating the luminosity function.

\subsection{Spectral extraction}
\label{xspectra}

\subsubsection*{XMM-XXL-N}
The XXL survey \citep{pierreXXL} was built via a mosaic of multiple \XMM\ observations with a high degree of overlapping between observations. This means that a single X-ray source can be observed multiple times. To analyze the X-ray spectra in the XXL sample, we used all the pointings contributing to a given source position. In particular, for each \XMM\ observation and for each EPIC camera (PN, MOS1, MOS2) available, we extracted the source and background spectra, and their corresponding ancillary response files (ARFs) and response matrix files (RMFs). All observations were reprocessed using the \XMM\ Science Analysis System (SAS, version 19) with the most updated calibration up to the date of analysis. The spectra were extracted following the standard procedure outlined by the SAS documentation. We used the \texttt{eregionanalyse} SAS task in order to calculate optimal elliptical source extraction regions centred at the position of each source. The orientation and eccentricity of the ellipse is defined by the PSF at the given position in the detector, and the final area of the region is calculated to maximize the signal-to-noise ratio (SNR) of the spectrum. Background spectra were extracted in circular regions of 30~arcsec centred at positions 1.5~arcmin away from the source. The exact position of the background region was selected to maximise the number of good pixels in the region (after masking areas outside the detector, bad pixels and other nearby detected sources) and to be as close as possible to the detector column of the source. For each source, all available spectra are not co-added, but instead they are fitted at once in the spectral fitting procedure outlined below in Sect.~\ref{xfit}.

\subsubsection*{CCLS}
We used CIAO\,4.13 \citep{Fruscione2006} for extracting the \Chandra\ X-ray spectra for sources in the CCLS field, following the \citet{Laloux2023} methodology to optimise the SNR for each spectrum. A source extraction region is defined as a circular area at the position of the source. Different radii are tested, each corresponding to an encircled energy fraction (EEF) radius from 50\% to 95\%. The background region is defined as an annulus of width 17.5\,arcsec with an inner radius 2.5\,arcsec larger than the source region. Contributions from nearby sources were removed from the background region. The SNR is calculated for each EEF value from the number of counts in both regions, and the region with the maximum SNR was selected. For each source, the spectra from all individual \Chandra/ACIS-I observations were extracted with the \texttt{specextract} task and then combined using the \texttt{combine\_spectra} task that also combines the ARF and RMF matrices.

\subsubsection*{CDF-S/N}
For the sources in the CDF-S/N fields, we used the X-ray spectra from \citet{Vito2018} who extracted the \Chandra\ spectra using the CIAO \texttt{ACISextract} package \citep{Broos2010} following a similar procedure to \citet{Luo2017} and \citet{Xue2016}. For the additional high-z sources identified in this work, we extracted the spectra following the procedure outlined for the CCLS field.

\subsection{Spectral analysis}
\label{xfit}

We fitted the X-ray spectra of the sources in our high-z sample using Sherpa 4.14.0 \citep{Freeman2001,sherpa414} and BXA 4.1.1. BXA is a Python package that connects a nested-sampling Monte Carlo algorithm \citep{Skilling2004} as implemented in \texttt{UltraNest} \citep{Buchner2021}, with the fitting software Sherpa, allowing a fully Bayesian approach for the X-ray spectral analysis. In this approach, the estimated background emission is not subtracted, but modelled using the method presented in \citet{Simmonds2018}, who did a principal component analysis (PCA) of archival data for different X-ray missions. We applied the \Chandra/\XMM\ PCA models for fitting the background spectrum using a standard Cash minimization with a Levenberg-Marquardt algorithm. Once a reasonable fit is obtained, the model parameters are fixed and the normalization re-scaled to the area and effective exposure time of the source extraction region. This background model is incorporated into the total model used for fitting the source spectrum. To take into account small discrepancies between the background spectrum in the extraction region and the background in the source region, we treated the normalization of the background component in the source spectrum modelling as a free parameter, using a log-normal prior with mean equal to the scaled normalization value and a dispersion of fifty per cent the value of the scaled normalization.

We modelled the source emission using \texttt{UXClumpy} \citep{Buchner2019}, corrected with a multiplicative absorption component \citep[\texttt{TBABS},][]{Wilms2000} to take into account the Galactic \nh\ along the line-of-sight of the source.\footnote{We use \texttt{gdpyc} \citep{gdpyc} to calculate the Galactic \nh\ from the Leiden/Argentine/Bonn (LAB) survey \citep{Kalberla2005}.} \texttt{UXClumpy} gives the reprocessed X-ray emission of the central AGN engine (a power-law with an exponential cut-off) by a clumpy torus, where individual high-density gas clouds are distributed following a toroidal geometry. It includes three components: the transmitted emission through the absorbing clouds, the reflected emission (including fluorescent lines) and the warm back-scattered emission containing mostly the incident power-law from unobscured sight-lines. This model is suitable for both type 1 and type 2 AGNs, since low inclination angles allow for a direct view of the central, unabsorbed emission. 

Our model has five free parameters: the $\log$\nh~along the line-of-sight of the observer, the photon index of the direct X-ray emission ($\Gamma$), the inclination angle of the torus ($\theta$) with respect to the line-of-sight of the observer, the logarithm of the normalisations for the direct emission and the log of the fraction of the scattering component with respect to the direct emission ($\log f_s$). We used flat priors for $\log\mnh$~(with limits between 20-26), $\theta$ (0-90 degrees), $\log f_s$ (between -5 and -1.3, i.e. up to a 5 per cent contribution for the scattered emission) and the normalisation. For $\Gamma$, we assumed a Gaussian prior with mean 1.95 and standard deviation 0.15, following the typical distribution of $\Gamma$ expected for AGNs \citep[e.g.][]{Nandra1994}. In the case of sources with photometric redshifts, we followed the procedure of \citet{Ruiz2021} and treated the redshift as an additional free parameter using as a prior the corresponding PDF given by the photo-z estimation software. We kept the remaining parameters of \texttt{UXClumpy} fixed during the fit, at the default values of the model. The energy range we used for the fit was $0.3-10\,\mathrm{keV}$ for \XMM\ spectra and $0.3-8\,\mathrm{keV}$ for \Chandra\ spectra. We evaluated the goodness of our spectral fits using the Cash-based test proposed by \citet{Kaastra2017}. For a detailed presentation of this method, as well as a discussion of the reliability of our results and the informational gain we obtained for the parameters, see Appendix~\ref{app_xfit}.

One of the advantages of using BXA is that the final result of the fitting process is the full posterior distribution of the free parameters of our model, conserving all possible degeneracies and correlations between the parameters. Errors for each parameter can be calculated from the posterior via marginalization. Derived quantities like observed fluxes or absorption corrected luminosities can be calculated using the full posterior, allowing us to correctly propagate the correlations between parameters and an accurate estimation of the errors of these derived quantities. In practical terms, the resulting BXA posteriors are equivalent to Monte Carlo Marko Chains and they can be analyzed using the tools already available in X-ray fitting software like Sherpa or XSPEC.

In Fig.~\ref{distributions}, we show the absorption column density (upper) and the absorption-corrected X-ray luminosity (lower) distributions of our final high-z sample. We plot the sum of the posterior probability distribution functions of individual sources coming directly from the spectral analysis along with their uncertainties. The uncertainties of the derived parameters besides on the photon statistics also depend on the overall spectral shape and on the accuracy of the adopted redshift measurements. For comparison, we also show the histograms when considering only the nominal values (mode of the PDF) with the highest probability of the spectral analysis. The most prominent difference is found in the number density in the $\rm \log N_H[cm^{-2}] = 20-23$ range. Using the sum of the posterior distributions, the number of sources in the first bins are of almost equal probability. This is something expected if we consider that at these high redshifts it is not possible to constrain the absorption below $\rm \log N_H[cm^{-2}]=23$ due to the shifted spectrum to higher energies. In contrast, if one uses the mode of the posterior distribution for each source would loose information, and hence, this would result in wrong conclusions. Furthermore, in the last bin of the $N_{\rm H}$ distribution there is an excess of sources when using the full posteriors that can be attributed to the tails at high values.

Figure~\ref{z-lx} presents the distribution of the high-z sources (selected in the different X-ray surveys) on the intrinsic X-ray luminosity (absorption-corrected) versus redshift plane (left panel) and on the absorption column density versus intrinsic X-ray luminosity plane (right). Each data point corresponds to the most probable values of the 3-dimensional X-ray luminosity, column density and redshift probability distribution function of individual sources inferred from the X-ray spectral fits. This figure demonstrates the necessity of combining various surveys with a wide range of areas and depths that are complementary to each other to map with the highest possible completeness the full AGN population.

\section{X-ray luminosity and absorption functions}
\label{xlaf}

\subsection{Survey area function}\label{areaa}

\begin{figure}
\center
\begin{tabular}{c}
    \includegraphics[width=0.47\textwidth]{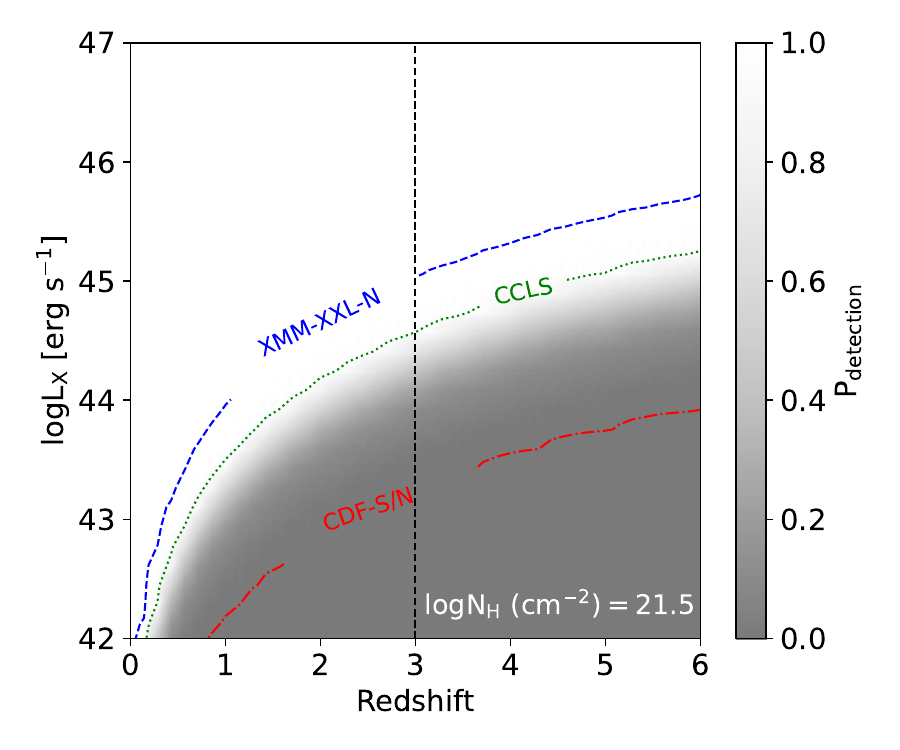} \\
    \includegraphics[width=0.47\textwidth]{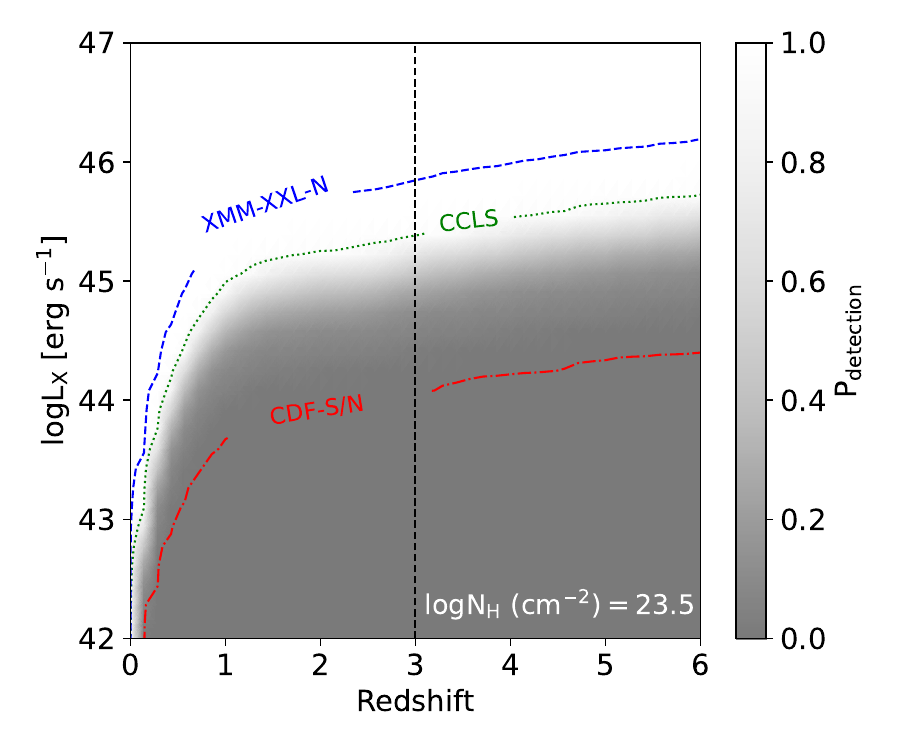} 
\end{tabular}
\caption{Sensitivity maps of the total area as a function of redshift and intrinsic X-ray luminosity for a source with intrinsic column density of $\log \mnh=21.5$ (upper) and $\log \mnh=23.5$ (lower). For comparison, we show the sensitivity maps at 99\% within the CDFs, CCLS and XMM-XXL-N independently.}\label{areasurvey}
\end{figure}

The sensitivity of X-ray surveys is not homogeneous, but decreases strongly toward faint fluxes (see Fig.~\ref{area_curves}). This introduces complex observational biases against sources with high $\mnh$, high redshift or lower X-ray luminosities (e.g. highly absorbed sources show lower X-ray fluxes so they are less likely to be detected). Such biases must be quantified if we want an accurate estimation of the luminosity and absorption functions for the intrinsic population of the high-z AGNs.

Using the {\sc UXClumpy} model, we calculated the expected flux as a function of $z$, $\LX$ and $\mnh$ assuming a constant value for the photon index, $\Gamma=1.95$ \citep{Nandra1994}. The parameter space used here was within $\log \LX = 42 -47$, $\log \mnh=20-26$ and $z=3-6$ with 50 bins for each parameter.\footnote{Along the paper, numerical values for the logarithm of $\mnh$ and $\LX$ are always quoted in CGS units, i.e. $\log(\mnh/ \mathrm{cm^{-2}})$ and $\log(\LX/ \mathrm{erg\,s^{-1}})$.} The inclination angle was fixed to $i=60\deg$ and the torus opening to $\sigma=30\deg$ that is a mixture of type 1 ($\sigma=15\deg$) and type 2 ($\sigma=40\deg$) AGN \citep{Buchner2019}. Fixing these parameters to different values does not affect significantly the detection probability of the sources. Then, we converted the fluxes into the corresponding area covered by each field by convolving with the area curves shown in Fig.~\ref{area_curves}. By normalizing, we were able to obtain the probability of a source with specific column density and intrinsic luminosity being detected at a given redshift and survey. 

The upper (lower) panel in Fig.~\ref{areasurvey} shows the probability of detecting an unobscured (obscured) source with $\log \mnh=21.5$ (23.5) within the combined area of all fields (shaded region). For comparison, we show the sensitivity curves within the CDF-S/N (dashed-dotted line), CCLS (dotted line) and XMM-XXL-N (dashed line) fields individually. As expected, there is a rapid drop of the probability for sources at high redshifts and low luminosities. This trend is more obvious for the obscured population. For example, at a given redshift and luminosity, the probability of detecting a source with $\log \mnh=21.5$ is higher than  considering higher column densities. Furthermore, the efficiency of detecting sources with lower intrinsic column densities and luminosities is higher in fields with deeper observations.

\subsection{X-ray luminosity function}
\label{sec:xlf}
We define $\phi$ as the differential luminosity function of the AGN population in terms of $\rm \log L_X$, since our high-z sample spans a wide luminosity range. By definition, $\phi$ is the number of sources $N$ per comoving volume $V$ and per logarithmic interval $\log\LX$ as a function of redshift, $z$, and luminosity, $\LX$:
\begin{equation}
\phi(\LX,z) = \dv{\Phi(\LX,z)}{\log\LX} = \dfrac{\dd[2]{N}(\LX,z)}{\dd V \dd\log\LX}
\end{equation}
% \begin{equation}
% \phi (\log \LX,z)=\frac{d^2N}{dVdlogL_{X}}(logL_X,z)=\frac{d^2N}{dVdlogL_{X}}(logL_X,z).
% \end{equation}

We derived the analytical expression of the differential luminosity function by assuming a broken power-law, which has been found to describe the shape well in the local and the nearby Universe \citep{Maccacaro1983,Maccacaro1984,Barger2005}, and is defined as:
\begin{equation}
\dv{\Phi(\LX,z=0)}{\log\LX} =
A \times \left[\left(\dfrac{\LX}{L_*}\right)^{\gamma_1} + \left(\dfrac{\LX}{L_*}\right)^{\gamma_2}\right]^{-1},
\end{equation}
where $A$ is a normalization factor, $L_*$ is the characteristic luminosity break, while $\gamma_1$ and $\gamma_2$ are the slopes of the power-law before and after $L_*$, respectively \citep{Miyaji2000,Hasinger2005}. 

In addition, we introduced the evolution of the luminosity function with the redshift testing different models. In particular, we adopted the pure density evolution model \citep[PDE,][]{Schmidt1968} and the luminosity-dependent density evolution model \citep[LDDE,][]{Schmidt1983,Miyaji2000} that have been used extensively in the literature. Both can be expressed as:
\begin{equation}
\dv{\Phi(\LX,z)}{\log\LX} = 
\dv{\Phi(\LX,z=0)}{\log\LX} \times e(z)
\end{equation}
where $e(z)$ is the factor that characterizes the evolution with redshift and can be given as:
\begin{equation}
e(z) = \qty(\dfrac{1+z}{1+z_c})^\pden
\end{equation}
for the PDE model, while for the LDDE model there is an additional dependence on luminosity \citep{Vito2014,Georgakakis2015} such as:
\begin{equation}
e(z,\LX) = \qty(\dfrac{1+z}{1+z_c})^{\pden + \beta(\log \LX - 44)}
\end{equation}
The parameters $\pden$ and $\beta$ give the slope of the power law and the dependence on the luminosity, respectively. $z_c$ is the critical redshift fixed at $z_c=3$ \citep{Vito2014,Georgakakis2015}. The physical explanation of the PDE model is that the AGN population changes in numbers but its luminosity remains the same. Thus, only the normalization of the XLF varies with redshift. The LDDE models assumes further that the variation of the AGN number density depends on the luminosity. This results in the change of the shape of the luminosity function. 

For visualization reasons, we also calculated the binned luminosity function of the high-z AGN by dividing the sample into luminosity, redshift and absorption column density bins. To construct the binned luminosity function, we used the \citet{Page2000} method that is an updated version of the $1/V_{max}$ method \citep{Schmidt1968,Avni1980}. Thus, the binned luminosity function in a given range of redshift, luminosity and Hydrogen column density can be estimated by:
\begin{equation}
\phi(\LX,z,\mnh) = \dfrac{\langle N \rangle}{\iiint\Omega(\LX,z,\mnh) \dv{V}{z}\, \dd\log\LX\,\dd\log\mnh~\dd z} ,
\end{equation}
where $\langle N \rangle$ is the number of sources in the specific bin, $\dv*{V}{z}$ is the differential comoving volume, and $\Omega$ is the survey sensitivity function defined in Sect.~\ref{areaa}.

\subsection{Absorption function}
\label{absfun}
In Sect.~\ref{xrayproperties}, we derived the X-ray spectral properties of the individual high-z sources and presented the $\log \mnh$ distribution without taking into account observational biases (Fig.~\ref{distributions}). Here, we formulate the intrinsic absorption distribution function of AGN taking into account the redshift and luminosity dependencies. Following the methodology of \citet{Ueda2003,Ueda2014,klod2022}, we model the \nh\ function, $\fabs$, by combining flat-step functions for different \nh\ bins. 

As mentioned in Sect.~\ref{xfit}, our analysis of the \XMM\ and \Chandra\ spectra was limited in the low-energy range to 0.3~keV in the observer frame, which corresponds to a limit in the rest-frame of our sample of about $1.2-2$~keV. This means that we are not able to reliably constrain values of $\log \mnh$ below $\sim22.5-23$, since absorption below these column densities only affects the X-ray spectrum below our observing energy range. Therefore, our absorption function is split into three \nh\ bins, as follows:
\begin{equation}
\fabs(z, \LX, \mnh) = 
\begin{cases}
\dfrac{1}{3} - \dfrac{\varepsilon}{3(1 + \varepsilon)} \psi(z,\LX) & [20 \leq \log \mnh < 23] \\
\dfrac{\varepsilon}{1 + \varepsilon} \psi(z,\LX) & [23 \leq \log \mnh < 24]\\
\dfrac{\fctk}{2}~\psi(z,\LX) & [24 \leq \log \mnh < 26]
\end{cases}
\end{equation}
Following the definitions of \citet{klod2022}, the parameter $\varepsilon$ is the ratio of sources with $23\leq\log\mnh\leq24$ to those with $22\leq\log\mnh\leq23$, while $\fctk$ gives the relative ratio between Compton-Thick (CTK, $24\leq\log\mnh\leq26$) sources and absorbed Compton-Thin (CTN, $22\leq\log\mnh\leq24$) objects. The term $\psi(z,\LX)$ corresponds to the fraction of absorbed CTN AGNs to the total number of AGN.

To compare with previous works \citep{Ueda2014,klod2022}, we use the following complex notation system where the redshift and luminosity dependence is contained in $\psi$, and it is parameterized using a linear dependence for $\log \LX$:
\begin{equation}\label{beta}
\psi(z,\LX)=\min(\psi_{\rm max}, \max(\psi_{43.75}(z) - c(\log \LX - 43.75), \psi_{\rm min})),
\end{equation}
where we used $\psi_{\rm min}=0$ and $\psi_{\rm max}=0.99$. The parameter $c$ controls the luminosity dependence and $\psi_{43.75}(z)$ represents the absorption function of AGN at $\log \LX=43.75$ for a given redshift. $\psi_{43.75}(z)$ is well-constrain for $z<2$ \citep{Ueda2014}, while above this redshift usually it is considered as a constant \citep[$2\leq z <3$,][]{klod2022}. In this work, we define $\psi_{43.75}(z)$ for sources with $z\geq3$, such as:
\begin{equation}
\psi_{43.75}(z \geq 3) = \psi_3 \times (1+z)^{a_2},
\end{equation}
where $\psi_3$ is the absorption function at $\log \LX=43.75$ and $z = 3$, with $a_2$ being the evolution index. Both are free parameters to be determined from the analysis in the next section.

\section{Results}\label{section_fit}

\subsection{Fit and parameter estimations}\label{fit}
\begin{figure*}
\center
    \includegraphics[width=1\textwidth]{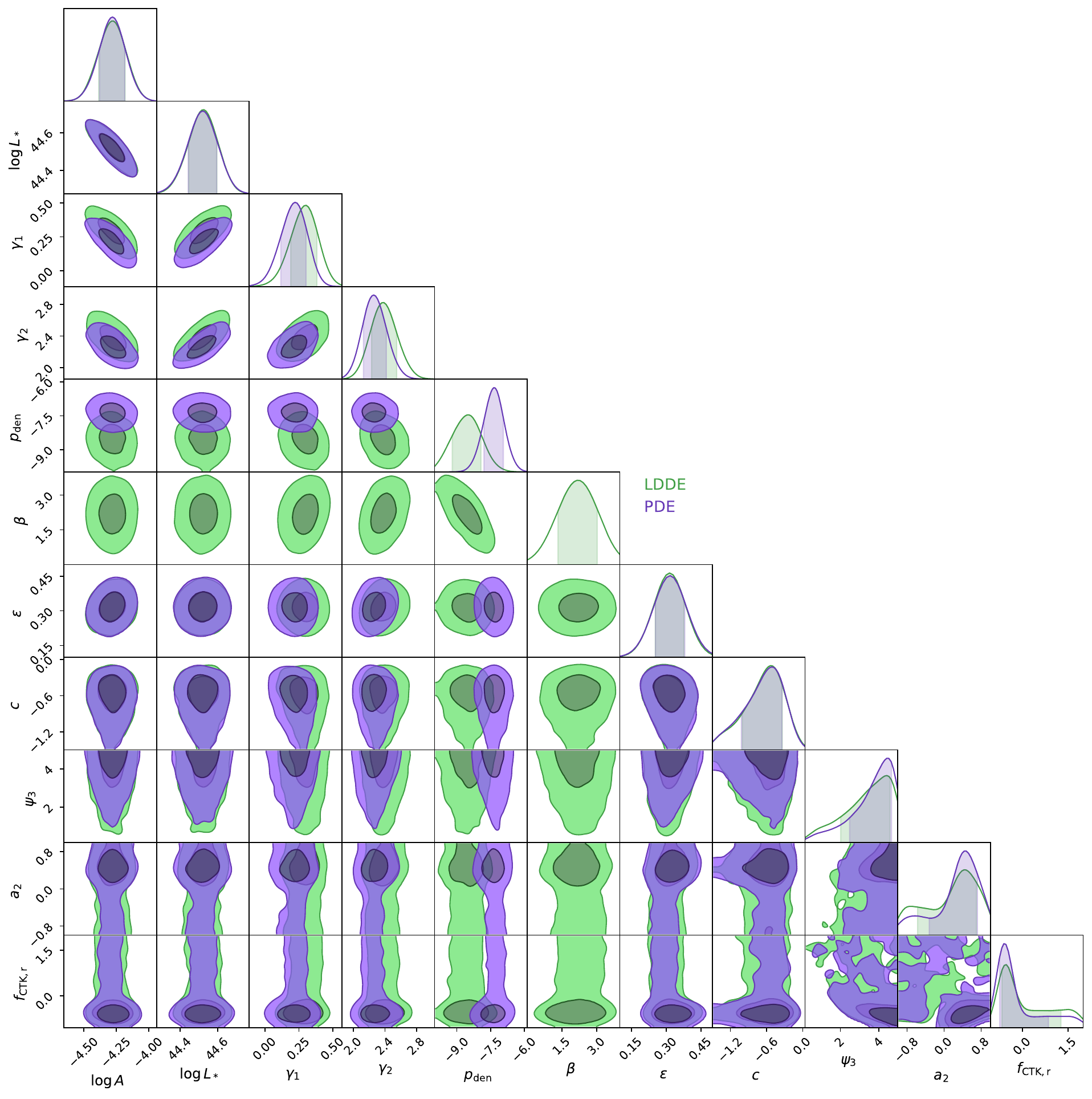} 
\caption{One-dimensional (diagonal panels) and two-dimensional marginal posterior distributions for the PDE (purple) and LDDE (green) model parameters. The shaded areas in the 2D posterior distributions correspond to $1\sigma$ and $2\sigma$ confidence levels (2D values, i.e. 39\% and 86\% respectively). The shaded areas for the 1D posteriors correspond to $1\sigma$ confidence level.}\label{posteriorsA}
\end{figure*}

\begin{figure}
\centering
\includegraphics[width=\hsize]{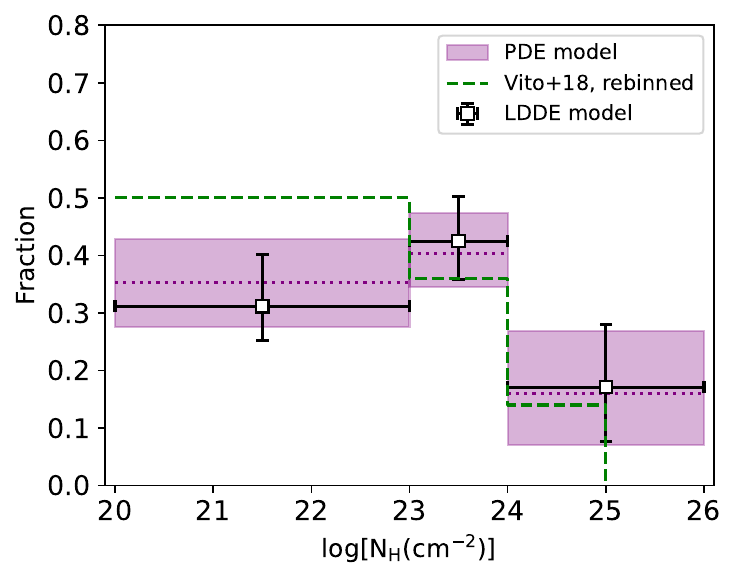}  
\caption{The best-fitting intrinsic absorption function in the redshift range $3\leq z\leq 6$. The boxes (points) show the credible regions that correspond to 1$\sigma$ confidence intervals of the posterior probabilities when using the PDE (LDDE) model. The dotted lines (points center) correspond to the median values of the absorption function. For reference, we show the results of \citet{Vito2018} re-scaled to the bins of our analysis (dashed line). The solid line presents their original absorption function. We note here that the latter is normalised between $\log N_{\rm H}=20-25$.}\label{plot_fabs}
\end{figure}

\begin{figure}
\centering
\includegraphics[width=0.45\textwidth]{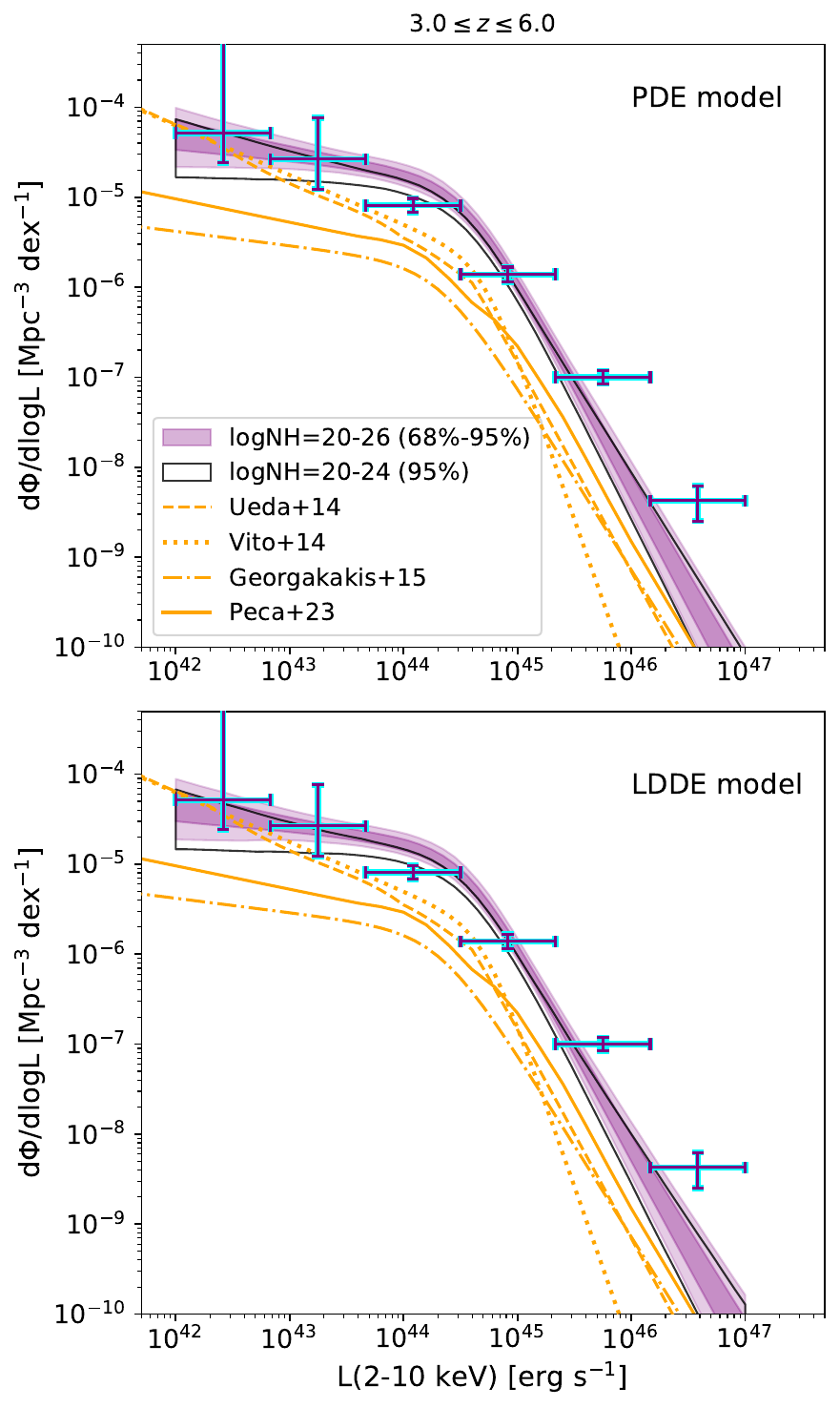}  
\caption{X-ray luminosity function in the redshift range $3.0\leq z \leq 6.0$ for the PDE (top) and LDDE model (bottom). The shaded regions represent the 68\% and 95\% confidence intervals of the XLF when integrating over $20\leq \log N_{\rm H} \leq 26$, while the black lines show the XLF when integrating over $20\leq \log N_{\rm H} \leq 24$. The points show the binned luminosity function. For comparison, we show the XLFs derived by previous X-ray studies in the column density range $\log N_{\rm H} = 20-24$. The best-fitting models of these studies are evaluated at the mean redshift of each bin.}\label{plot_XLF_zbins}
\end{figure}

\begin{table}
\caption{Best-fit parameters of the XLF and absorption function in the cases of the LDDE and PDE models.}   \label{tableLDDE}      % is used to refer this table in the text
\centering                                      % used for centering table
\begin{tabular}{l  c  r  r }          % centered columns (4 columns)
\hline\hline                        % inserts double horizontal lines

Parameter  & Prior & \multicolumn{2}{c}{Best value} \\

\hline                                   % inserts single horizontal line
       & XLF &  LDDE & PDE \\
        \hline                                   % inserts single horizontal line
\rule{0pt}{2ex} 
        $\log A$         & -7, -3 & $-4.28^{+0.09}_{-0.10}$ &   $-4.28^{+0.09}_{-0.09}$\\ \rule{0pt}{3ex}  
        $\log L_*$       & 42,  46& $44.52^{+0.07}_{-0.07}$ &   $44.52^{+0.07}_{-0.07}$ \\ \rule{0pt}{3ex}
        $\gamma_1$      & -2,  2&   $0.29^{+0.08}_{-0.10}$      &$0.21^{+0.08}_{-0.10}$\\ \rule{0pt}{3ex}
        $\gamma_2$      & 1,  6&    $2.38^{+0.15}_{-0.14}$      &$2.23^{+0.14}_{-0.13}$\\ \rule{0pt}{3ex}
        $\pden$      & -10, -3&      $-8.53^{+0.58}_{-0.65}$     &$-7.35^{+0.39}_{-0.42}$\\ \rule{0pt}{3ex}
        $\beta$        & -3, 5&     $2.18^{+0.79}_{-0.81}$      &--\\ 
        
        \hline                                 

        \multicolumn{4}{c}{Absorption function} \\ 
        \hline                                   % inserts single horizontal line
\rule{0pt}{2ex} 
        $\varepsilon$        & 0.1, 10& $2.041^{+0.24}_{-0.24}$ & $2.09^{+0.23}_{-0.23}$\\  \rule{0pt}{3ex}
        $a_2$            & -2, 2& $-0.28^{+1.57}_{-0.37}$ & $-0.50^{+1.37}_{-0.21}$\\ \rule{0pt}{3ex}
        $\psi_3$        & 0, 5& $3.66^{+0.95}_{-1.62}$ & $3.99^{+0.74}_{-1.45}$\\  \rule{0pt}{3ex}
        $c$            & -1, 1& $0.28^{+0.42}_{-0.87}$ & $0.39^{+0.32}_{-0.72}$\\ \rule{0pt}{3ex}  
        $\fctk$    & 0.03, 10 & $0.25^{+0.14}_{-0.22}$ &  $0.25^{+0.14}_{-0.23}$ \\
\hline                                             %inserts single line
\end{tabular}
\tablefoot{The normalisation $A$ and the break luminosity $L_*$ are given in units of $Mpc^{-3}$ and $erg~s^{-1}$, respectively.}
\end{table}

In this work, we used Bayesian inference to estimate the parametric form of the X-ray luminosity function and the absorption function simultaneously. Given a data-set of $n$ observations, $D = \{d_i; i=1,...,n\}$, and a model for the X-ray luminosity function defined by a set of parameters $\vb*{\Theta}$, according to the Bayes' theorem:
\begin{equation}
P(\vb*{\Theta} | D) = \dfrac{P(D|\vb*{\Theta}) P(\vb*{\Theta})}{P(D)},
\end{equation}
where $P(\vb*{\Theta} | D)$, the posterior probability, is the probability of the selected model given the observational data; $\La = P(D | \vb*{\Theta})$, the likelihood, is the probability of obtaining the observational data given the model; $P(\vb*{\Theta})$, the prior, is the \textit{a priori} probability for the parameters of the model, and $P(D) = \int P(\vb*{\Theta} | D) \dd\vb*{\Theta}$ is the evidence of the model.

To derive the posterior probability distribution of the model parameters, we used the nested-sampling Monte Carlo algorithm MLFriends \citep{Buchner2016,Buchner2017}, implemented in the \texttt{UltraNest} package. Nested sampling algorithms allow tracing the posterior distribution of the model, given a data set, while at the same time calculating the Bayesian evidence. A direct estimate of the evidence is extremely useful for the comparison of the different XLF models via Bayes factors (see Sect.~\ref{results}). Moreover, this Bayesian approach allows for a rigorous treatment of the uncertainties in the X-ray properties and photometric redshifts of the sources obtained in our X-ray analysis. During the inference process, we assumed flat priors for the model parameters, either uniform or log-uniform, that span a reasonably broad range of the parameter space according to previous studies in the literature \citep{Vito2014, klod2022}. In Table~\ref{tableLDDE} we provide the minimum and maximum values allowed in the flat priors we used for the parameters of the XLF models.\footnote{A Python implementation of the presented methodology used to derive the X-ray luminosity and absorption functions is available online at \url{https://github.com/ruizca/xlaf}.}

\citet{Loredo2004} proposed that the likelihood of observing a given data set can be constructed as the product of the probabilities of observing each individual source times the probability of not detecting any other source. Following the detailed derivation of \citealt{Buchner2015} \citep[see also][]{Aird2015,Georgakakis2015}, the log-likelihood of this process can be written as:
\begin{multline}\label{eq:xlflikelihood}
\ln \La(\{d_i\} | \vb*{\Theta}) = \\
\shoveleft-\lambda + \sum_i \ln\iiint P_i(\LX, z, \mnh | \vb*{\Theta})~\dv{V}{z}\dlog\mnh~\dlog\LX~\dd z.
\end{multline}
The parameter $\lambda$ is the expected number of observed sources for a Poisson process, given an XLF model with parameters $\vb*{\Theta}$:
\begin{equation}\label{eq:expectedsources}
\lambda = \iiint \phi_\mathrm{abs}(\LX, z, \mnh | \vb*{\Theta}) \Omega(\LX, z, \mnh)~\dv{V}{z}\dlog\mnh~\dlog\LX~\dd z,
\end{equation}
where $\Omega$ is the survey sensitivity function calculated in Sect.~\ref{areaa} and $\phi_\mathrm{abs} = \phi \times f_\mathrm{abs}$. $\phi$ and $f_\mathrm{abs}$ are the luminosity and absorption functions defined by the parametrizations presented in Sects.~\ref{sec:xlf} and~\ref{absfun}, respectively.

The parameter $P_i$ in Eq.~\ref{eq:xlflikelihood} is given by:
\begin{equation}
P_i(\LX, z, \mnh | \vb*{\Theta}) = p(d_i | \LX, z, \mnh)~\phi_\mathrm{abs}(\LX, z, \mnh | \vb*{\Theta})~\Omega(\LX, z, \mnh).
\end{equation}
where $p(d_i | \LX, z, \mnh)$ is the probability of the source $i$ being at redshift $z$ with $\mnh, \LX$ X-ray properties. This probability is given by the posterior probability distributions we obtained during our X-ray spectral analysis (Sect.~\ref{xrayproperties}). We have included in this term the sensitivity function of the survey $\Omega$. While, according to \citet{Loredo2004}, this is formally incorrect, this term takes into account the loss of information due to the different methods used for X-ray source detection and for the X-ray spectral analysis \citep[see Appendix~A of][for a detailed discussion of this issue]{Buchner2015}. The integral involving $P_i$ in Eq.~\ref{eq:xlflikelihood} can be calculated using an importance sampling Monte Carlo integration technique \citep{Kloek1978,Press1992}.

The integration limits used in Eqs.~(\ref{eq:xlflikelihood}) and~(\ref{eq:expectedsources}) are [3,6], [42,47] and [20,26] for the parameters $z, ~\log \LX$ and $\log \mnh$, respectively. Overall, our combined parametrization of the luminosity and the absorption functions have eleven free parameters when we use the LDDE model for XLF, and ten parameters for the PDE model. In Table~\ref{tableLDDE} we report the best-fit parameter estimations with their uncertainties of the X-ray luminosity function and the absorption function derived adopting the LDDE and the PDE models. Figure~\ref{posteriorsA} presents the one-dimensional (diagonal panels) and two-dimensional marginal posterior distributions for the PDE (purple) and LDDE (green) model parameters. As can be seen from the one-dimensional plots, only two parameters cannot be fully constrained: $\fctk$ and $a_2$. Even though these parameters are well-constrained at $1\sigma$, the $2\sigma$ contours are not closed.

Figure~\ref{plot_fabs} shows the best-fitting intrinsic absorption function in the redshift range of our analysis. The boxes (points and errors) show the credible regions that correspond to 1$\sigma$ confidence interval of the posterior probabilities for the PDE (LDDE) model. As can be seen further from Table~\ref{tableLDDE}, the absorption function parameters estimations are similar for both XLF models. We compare the intrinsic absorption function with the one derived in \citet{Vito2018}. In order to do a proper comparison, we integrated their absorption function over the binning of our analysis (dashed lines). The fraction of heavily obscured sources ($\log N_{\rm H} \geq 23$) agrees with our results within the uncertainties, even though their average values appear lower. Their derived fraction of sources with $20 \leq\log N_{\rm H} \leq 23$ is about 50\% that is 15\% higher than in our case. Perhaps this is due to the fact that in their analysis the absorption function is normalised between $\log N_{\rm H}=20-25$ instead. Furthermore, we followed a different approach to compute the obscuration incompleteness that affect strongly the results.

\begin{figure*}
\center
    \includegraphics[width=1\textwidth]{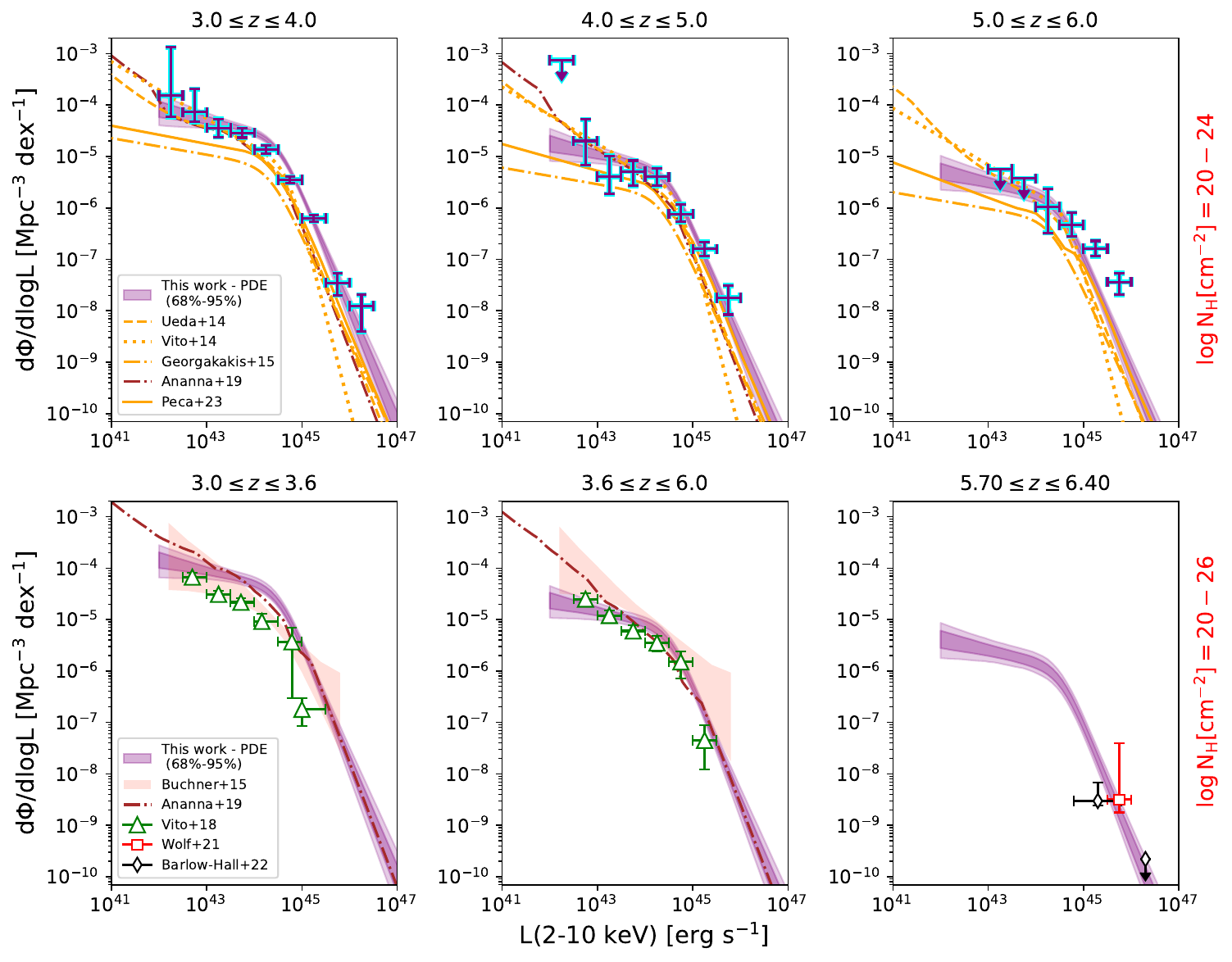}
\caption{The best-fitting PDE model in several redshift bins computed by integrating the XLF over redshift and column density. The purple shaded regions represent the 68\% and 95\% confidence intervals of the model, while the purple data points indicate the binned luminosity function. We compare our results with the parametric forms of the XLF derived by previous X-ray studies \citep{Ueda2014,Vito2014,Georgakakis2015,Peca2023} in the column density interval of $\log N_{\rm H} = 20-24$ (upper panels). Furthermore, we present our XLF integrating over $\log N_{\rm H} = 20-26$ (lower panels) to the results of \citep{Buchner2015,Vito2018,Wolf2021,BarlowHall2023}. The XLF of \citet{Buchner2015} in the first two panels are derived in the redshift range $3.2 \leq z \leq 4.0$ and $4.0 \leq z \leq 7.0$, respectively. Our PDE model at the last panel is extrapolated at z=6.05. The brown dashed-dotted lines show the predicted XLF derived by \citet{Ananna2019}.}
\label{xlf_full}
\end{figure*}

In Fig.~\ref{plot_XLF_zbins}, we plot the total X-ray luminosity function at the full redshift range of our analysis ($3 \leq z \leq 6$) with the 1$\sigma$ and 2$\sigma$ uncertainties for the PDE (top) and LDDE (bottom) models. The shaded (empty) regions correspond to the XLF that includes any absorption level between $\log N_{\rm H} = 20-26$ ($\log N_{\rm H} = 20-24$). We over-plot the binned luminosity function ($\log N_{\rm H} = 20-26$). The binned data appear to overestimate the XLF at high luminosities compared to the models. This comes from the  large uncertainties at $z>5$ and high column densities (as can be seen in the next section at Fig.~\ref{xlf_full}). For reference, we also show the results of previous X-ray studies \citep{Ueda2014,Vito2014,Georgakakis2015,Peca2023} at $\log N_{\rm H} = 20-24$. In the following section, we discuss further their similarities and differences with our results.

To quantify which of the two XLF models provides better fit to the data in the redshift range $3 \leq z \leq 6$, we used the information criteria: the Akaike Information Criterion \citep[AIC,][]{Akaike1974} and the Bayesian Information criterion \citep[BIC,][]{Schwarz1978} that have been used widely in the literature \citep[e.g., ][]{Fotopoulou2016a,Pouliasis2020}. They take into account the complexity of the models in addition to the goodness of fit and prefer models with less parameters. The best model according to these is the LDDE model. However, taking the difference of the two models, we found $DAIC=4.5$ and $DBIC=2.6$ that suggest that we may not exclude the possibility of the PDE model to be the correct one. Furthermore, we used a Bayesian model comparison. In particular, for each model we used the Bayes factor (the ratio between the evidences) of each model compared to the one with the highest evidence. It is possible to provide a measure of the weight of the information that is included in the data in a favor of one model against the other. Thus, according the interpretation of the Bayes factor, models that have high values (>1) are rejected. In our analysis, the LDDE model has the highest evidence. The Bayes factor of the PDE model is $\Delta\log\zeta=0.5$. With the Bayesian model comparison, we find that the two PDE and LDDE models represent the data equally well. In the following sections, we restrict ourselves to using only the PDE model as it has a lower number of parameters. However, we have to keep in mind that the LDDE model may represent equally well the observational data.

%The boxes (points) show the credible regions that correspond to 1$\sigma$ and 2$\sigma$ confidence intervals of the posterior probabilities when using the PDE (LDDE) model. For reference, we show the results of \citet{Vito2018} rebinned (dashed line). The solid line presents their original absorption function. We note here that the latter is normalised between $\log N_H=20-25$.

%Also, we found no luminosity dependence on the absorption function ($c = 0.01$) in both cases. 

%--------------------------------------------------------------------
\section{Discussion}\label{results}

\begin{figure*}
\center
\includegraphics[width=\hsize]{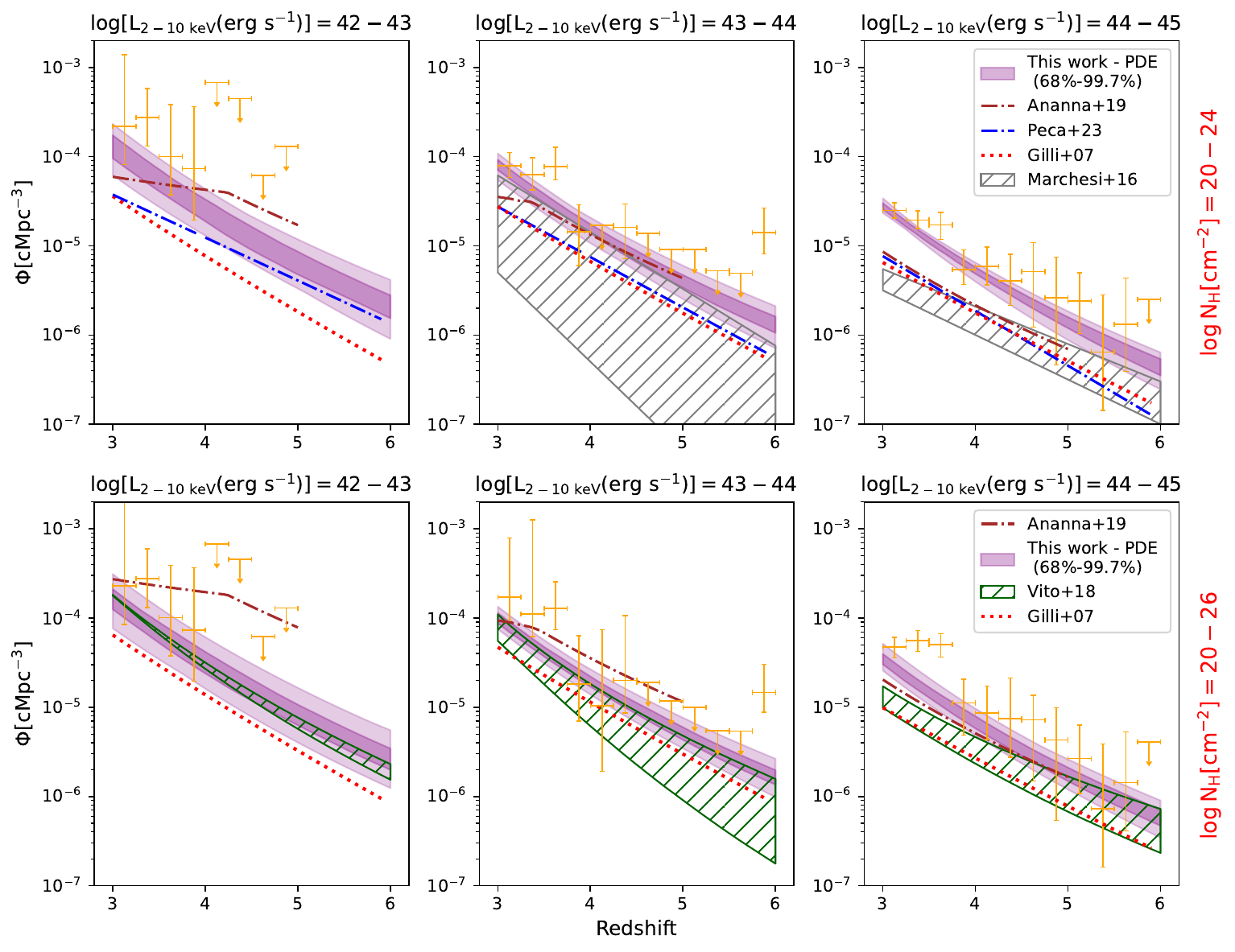} 
\caption{Redshift evolution of the AGN space density in three luminosity for $\log N_{\rm H} = 20-24$ (upper panels) and $\log N_{\rm H} = 20-26$ (lower panels). The shaded regions represent the 68\% and 99.7\% confidence intervals of the best-fitting PDE model, while the points represent the binned luminosity function. For comparison, we over-plot the results of \citet{Marchesi2016highz}, \citet{Vito2018} and \citet{Peca2023}. The dotted red lines indicate the LDDE model predictions with an exponential decline (with a power-law decay) by \citet{Gilli2007}. The brown dashed-dotted lines show the predicted XLF derived by \citet{Ananna2019}.}\label{spacedensity1}
\end{figure*}

\begin{figure*}
\center
    \includegraphics[width=\hsize]{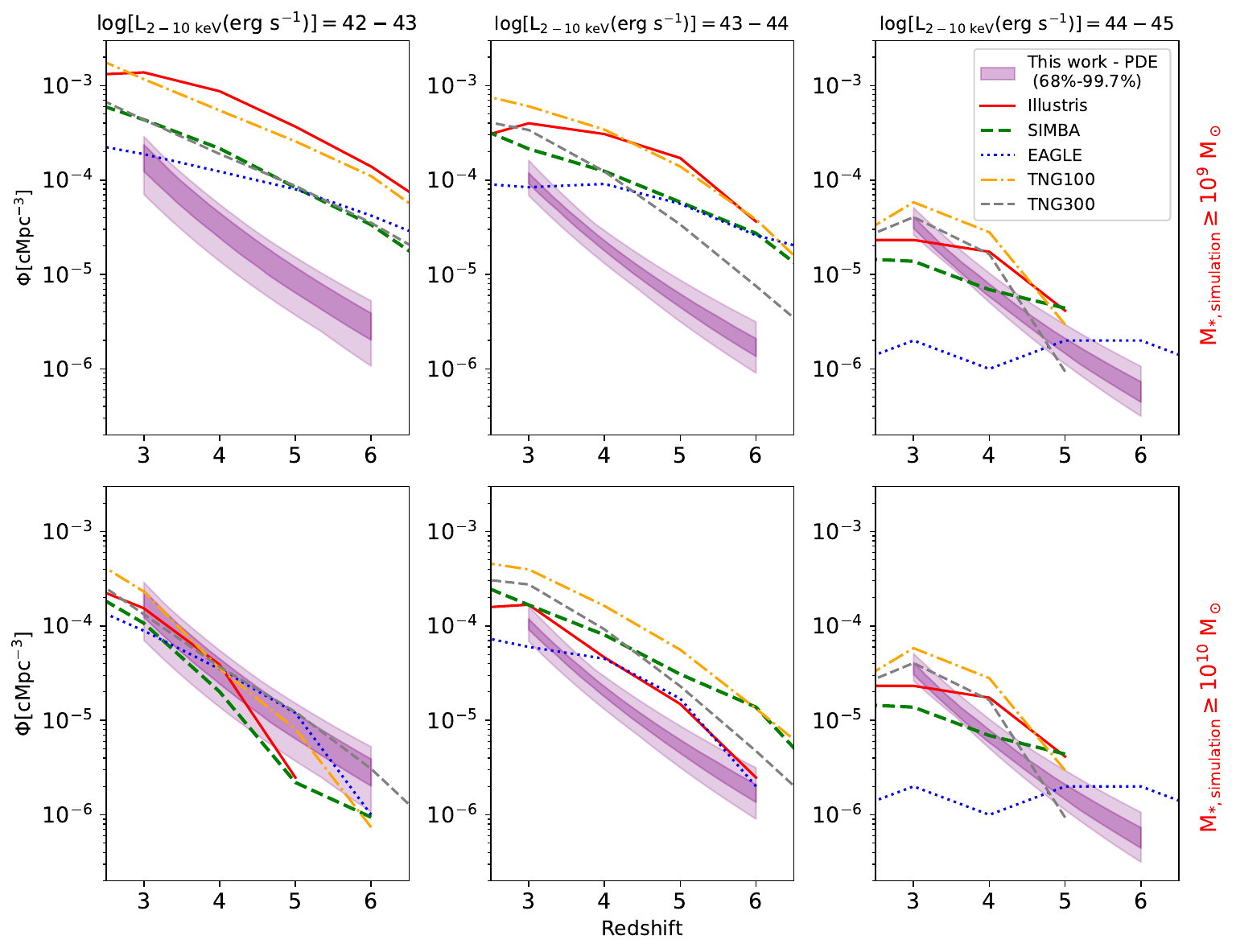} \\
\caption{Redshift evolution of the AGN space density in three luminosity bins as indicated on the top of the panels. The shaded regions represent the 68\% and 99.7\% confidence intervals of the best-fitting LDDE model. The lines represent the space densities derived from different cosmological simulations \citep{Habouzit2022}. The upper and lower panels correspond to systems with stellar mass higher than $M_*>10^9M_\sun$ and $M_*>10^{10}M_\sun$, respectively.}
\label{spacedensitySIM}
\end{figure*}

\subsection{X-ray luminosity function and comparison to previous constraints}\label{xlf}

In Fig.~\ref{xlf_full}, we plot our X-ray luminosity function in several redshift bins with the 1$\sigma$ and 2$\sigma$ uncertainties (shaded regions) for the PDE model. We over-plot the binned luminosity function and, for reference, we also show the results of the X-ray luminosity function derived in previous X-ray studies from the literature. In the upper panels, we compare our results with the parametric forms of the XLF derived by \citep{Ueda2014,Vito2014,Georgakakis2015,Peca2023} in the column density interval of $\log N_{\rm H} = 20-24$. \citet{Vito2014} and \citet{Georgakakis2015} selected 141 and $\sim$300 AGN, respectively, in the redshift range $3.0 \geq z \geq 5.0$ from various fields, while \citet{Ueda2014} gathered a sample of more than 4000 X-ray sources in the redshift range of $0.01<z<5$. More recently, \citet{Peca2023} used a large sample of luminous ($\log L_X \geq {44}$) AGN selected in the Stripe 82X field \citep{LaMassa2013a,LaMassa2013b,LaMassa2016} that covers an area of $31.3\deg^2$. We find a slightly higher number of sources compared to the aforementioned studies, especially at the knee of the luminosity function and towards the bright end. At lower luminosities ($\log L_X \leq {44}$), our results are consistent with the XLFs derived by \citet{Vito2014} and \citet{Ueda2014}.

In the first two lower panels of Fig.~\ref{xlf_full}, we present our XLF integrated over $\log N_{\rm H} = 20-26$ along with the works of \cite{Buchner2015} and \citet{Vito2018}. \citet{Buchner2015} studied a sample of about 2000 AGN selected from partly the same fields as in our case in the full redshift range $0.01<z<7$. Their XLF are derived in the redshift range $3.2 \leq z \leq 4.0$ and $4.0 \leq z \leq 7.0$, respectively. In the first redshift bin ($3.0\leq z \leq3.6$), our results agree with the work of \citet{Buchner2015} in the faint and bright end. However, in the break luminosity we find higher number of sources. The \citet{Vito2018} binned luminosity falls below our XLF. This is probably due to the fact that they have selected AGN only in the Chandra deep fields, missing this way the population of luminous ($\log L_X \geq 43$) AGN. In the redshift range $3.6\leq z \leq6.0$, our XLF is in a better agreement with \citet{Vito2018} within the uncertainties, while the number counts of \citet{Buchner2015} appear to over-predict the AGN density. In their study, they provided confidence regions at each redshift intervals instead of an analytical XLF form. Moreover, we compare our results to the XLF predictions derived by the \citet{Ananna2019} AGN population synthesis model in both column density bins. When we consider the full AGN population ($\log \mnh = 20-26$), our results are in a very good agreement at $\log L_X \geq \sim 44$, while they estimate higher values in the faint end. On the other hand, in the $\log \mnh = 20-24$ bin, their predicted XLF follows the one derived by \citet{Ueda2014} that is lower compared to our results. This probably comes from the fact that in their model the Compton-Thick fraction is as high as 50\% compared to the $\sim 17\%$ predicted in our analysis (Sect.~\ref{obscuration_section}).

To assess the agreement with recent observational constraints at the bright end of the function, we compare our results to the works of \citet{Wolf2021} and \citet{BarlowHall2023}. \citet{Wolf2021} identified a single source at $z=5.81$ in the eROSITA Final Equatorial-Depth Survey \citep[eFEDS,][]{Brunner2021} that covers an area of $140\deg^2$. \citet{BarlowHall2023} found a source at $z=6.31$ in the Extragalactic Serendipitous Swift Survey \citep[ExSeSS,][]{Delaney2023} catalogue. They provided as well an upper limit at higher luminosities. In Fig.~\ref{xlf_full} (bottom right), we plot our best-fit PDE model extrapolated at $z=6.05$ and integrated over $\log N_{\rm H} = 20-26$. The derived binned luminosity function of the aforementioned studies are consistent with our results within the uncertainties. This result suggests that our derived parameteric X-ray luminosity function may hold at even higher redshifts.

\subsection{Space density}\label{space}

In Fig.\ref{spacedensity1}, we plot the comoving space density of AGN versus redshift in three X-ray luminosity bins as indicated on the top of each panel. The parametric number density was calculated by integrating the XLF over X-ray luminosity and Hydrogen column density for the PDE model. The binned space density is represented with the data points. The upper and lower panels correspond to the space density when integrating over $\log N_{\rm H} = 20-24$ and $\log N_{\rm H} = 20-26$, respectively. We compare our results with previous X-ray studies \citep{Marchesi2016highz,Vito2018,Peca2023} and also with the predictions of the \citet{Ananna2019} model and that derived by \citet{Gilli2007} that shows a strong decline of the space desnity at high redshifts. Comparing our results with the space density derived by \citet{Marchesi2016highz}, we find a higher space density by a factor of 4-6. This difference can be ascribed to the fact that they did not take into account the obscuration incompleteness as it has been done in our analysis. We obtain similar results when we compare with the work of \citet{Peca2023}. Even though they have corrected for obscuration effects, it is possible that obscured sources are missing due to the shallower Stripe-82X ($\rm 8.7\times 10^{-16}~erg~cm^{-2}~s^{-1}$ in the soft band) and even the correction for the area is insufficient. Furthermore, in the redshift range $3<z<4$ the Stripe-82X does not include many sources, while above $z>4$ the space density is an extrapolation. Regarding the \citet{Vito2018} space density, in the lower luminosity bins, we find a comparable number of sources within the uncertainties, while for the most luminous AGN we predict a higher space density. This is probably due to the sparse sample at these luminosities in the Chandra deep fields, since the majority of the sources have $\log L_X\leq 43$. Finally, we find higher number density by a factor of 3-4 compared to the predictions of the LDDE model with an exponential decline by \citet{Gilli2007}. Regarding the \citet{Ananna2019} model, in the $\log \mnh = 20-24$ bin we find higher values at bright luminosities, while this difference diminishes going to the fainter end. For the full AGN population ($\log \mnh = 20-26$), the space density is in agreement within the uncertainties at bright luminosities. In the faintest bin, their model predicts higher values.

Figure~\ref{spacedensitySIM} (upper panels) compares our results with the predicted number densities from large-scale hydro-dynamical cosmological simulations \citep{Habouzit2021,Habouzit2022}. Taking into account the full AGN population predicted by simulations ($M_*>10^9M_\sun$), we find a large discrepancy with our results at lower luminosities. However, in \citet{Pouliasis2022b} we derived the stellar masses of the high-z X-ray sample selected in CCLS, XMM-XXL-N and the eFEDS fields and found that all the high-z X-ray AGN reside in galaxies with stellar masses above $10^{10}M_\sun$, with the majority being at $\geq 5 \times 10^{10}M_\sun$. Concerning the CDF-S/N sample, many studies also suggest that the majority of the high-z X-ray AGN have host galaxies with $M_\star>10^{10}M_\sun$ \citep{Santini2015,Yang2017,Circosta2019}. Thus, in order to compare properly the space density of the X-ray selected AGN to the one derived by the simulations, we have to restrict the sample of \citet{Habouzit2022} to sources with stellar masses of $M_*>10^{10}M_\odot$ (lower panels of Fig.~\ref{spacedensitySIM}).

In such a case, in the lowest luminosity bin ($\log \LX=42-43$), our results (both shape and normalisation) agree very well with all the predictions coming from all the different simulations. In the luminosity bin $\log \LX=43-44$, there is a discrepancy by a factor of around 10 compared to all simulations, even though the shapes are consistent. As mentioned above, the majority of the sources in \citet{Pouliasis2022b} have host galaxies with stellar masses greater than $5 \times 10^{10}M_\sun$. Hence, a higher stellar-mass cut in the simulations could diminish this difference. Concerning the most luminous AGN ($\log \LX=44-45$), our space density is consistent qualitatively with all the simulations but the EAGLE. Here, it is worth mentioning that the number densities from the simulations span at least half a dex, which highlights the large uncertainties in the number of AGN produced in the simulations and the complexity of modeling the black-hole and galaxy physics. Furthermore, these simulations do not include any correction for obscuration of any kind. A more thorough analysis where one would take into account all the selection biases and would use the same physical parameter's range between the simulations and the observed X-ray population is necessary to elaborate more. All the above point towards the fact that either the simulations over-predict the AGN in low-mass galaxies below $10^{10}M_\sun$ or the current X-ray surveys are not able to detect them. In the first case, this could be due to too massive black-holes in these simulated galaxies, too efficient BH accretion, or too weak supernova feedback (or even AGN feedback). The feedback explanation has been also suggested by \citet{Vito2018}. In Sect.~\ref{section_bhad}, we investigate further the second point (do X-rays miss the AGN population hosted by low-mass galaxies?) by comparing the BHAD derived in X-rays and in the mid-IR wavelengths through \textit{JWST} data.

\subsection{Intrinsic Obscuration fraction}\label{obscuration_section}

\begin{figure}
\center
   \begin{tabular}{c}
    \includegraphics[width=0.47\textwidth]{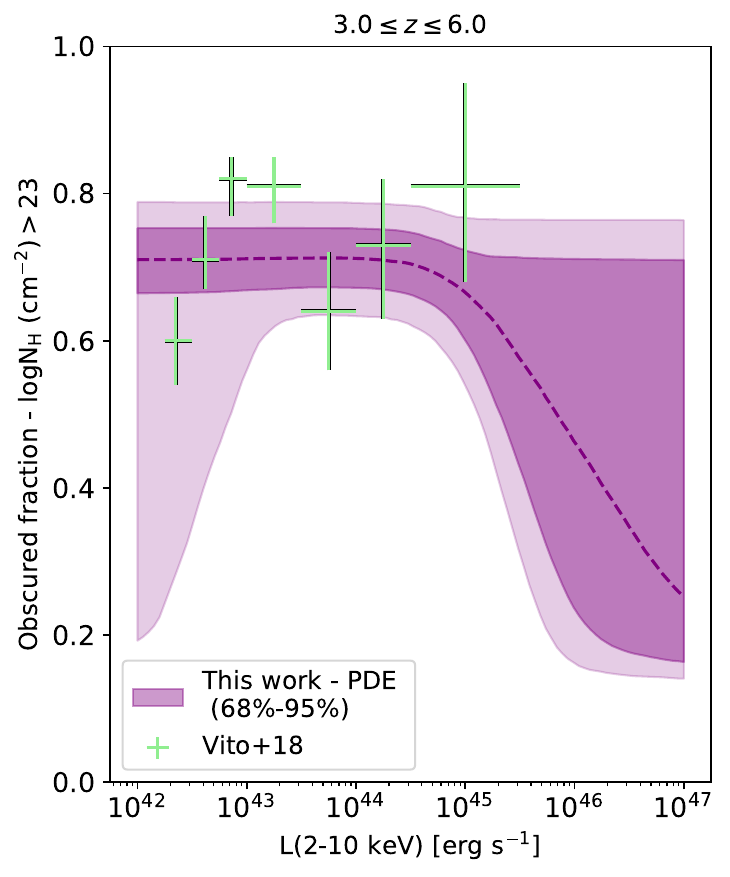} 
    \end{tabular}
\caption{Obscured fraction with $\log \mnh\geq23$ versus X-ray luminosity. Our results are shown with the shaded regions that represent the 68\% and 95\% confidence intervals of the best-fitting absorption function. For reference, we show the results of \citet{Vito2018}.}
\label{obscured_lx}
\end{figure}

The obscured fraction of the AGN population is important to understand how obscured and unobscured sources evolve through cosmic time and to study the accretion history of the super-massive black holes \citep{Hiroi2012,Iwasawa2012}. Heavily obscured sources by large column of gas ($\log \mnh\geq24$) play a key role both in the determination of the evolutionary models \citep{Alexander2012,Ricci2017} and in the population synthesis models to constrain the shape of the cosmic X-ray background \citep{Gilli2007,Ananna2019}. Due to the suppression of the spectrum at low energies because of obscuration, it is important to account properly for any observational biases to study this class of heavily obscured AGNs. Using the intrinsic \nh\ distribution derived from the minimization method described in Sect.~\ref{fit}, we find that the intrinsic fraction of Compton-Thick sources over the whole population is $F_{CTK}=0.17_{+0.07}^{-0.09}$ in the redshift range $3<z<6$. Comparing our results to the works of \citet{Burlon2011,Aird2015,Buchner2015,Masini2018,Georgantopoulos2019,Laloux2023} at lower redshifts, we find a constant Compton-Thick fraction from the local Universe up to redshift $z=6$. In a following paper (Pouliasis et al., in prep.), we will present exclusively the number density and properties of the Compton-Thick population.

In Sect.~\ref{xrayproperties}, we presented the observed distribution of the Hydrogen column density of the combined sample. The observed obscured fraction of sources with $\log \mnh>23$ is 28.7\% in the redshift range $3\leq z\leq6$ compared to the overall population when using the full \nh\ posterior distributions. Though, the intrinsic fraction of obscured AGN ($\log \mnh>23$) is $61_{-7}^{+8}\%$. As mentioned in the previous section, this fraction of obscured AGN concerns host galaxies mainly with $M_*>10^{10}M_\odot$. Among this obscured population, the relative fraction of Compton-Thick AGN with $\rm logN_H[cm^{-2}] \geq 24$ is $\rm F_{CTK,relative}=0.25_{-0.23}^{+0.14}$. In Fig.~\ref{obscured_lx}, we plot the obscured fraction with $\rm logN_H (cm^{-2})>23$ as a function of the intrinsic X-ray luminosity. We over-plot the fraction derived by \citet{Vito2018} in different luminosity bins. Despite the large uncertainties in the high-luminosity end, the results are consistent with each other indicating a flat fraction of $\sim$68\% across the hard X-ray luminosity. Similarly, we find a negligible dependence of the obscured fraction over cosmic time in the redshift range $3.0 \leq z \leq 6.0$ for the same luminosity bins. 

\begin{figure}
\center
   \begin{tabular}{c}
    \includegraphics[width=0.47\textwidth]{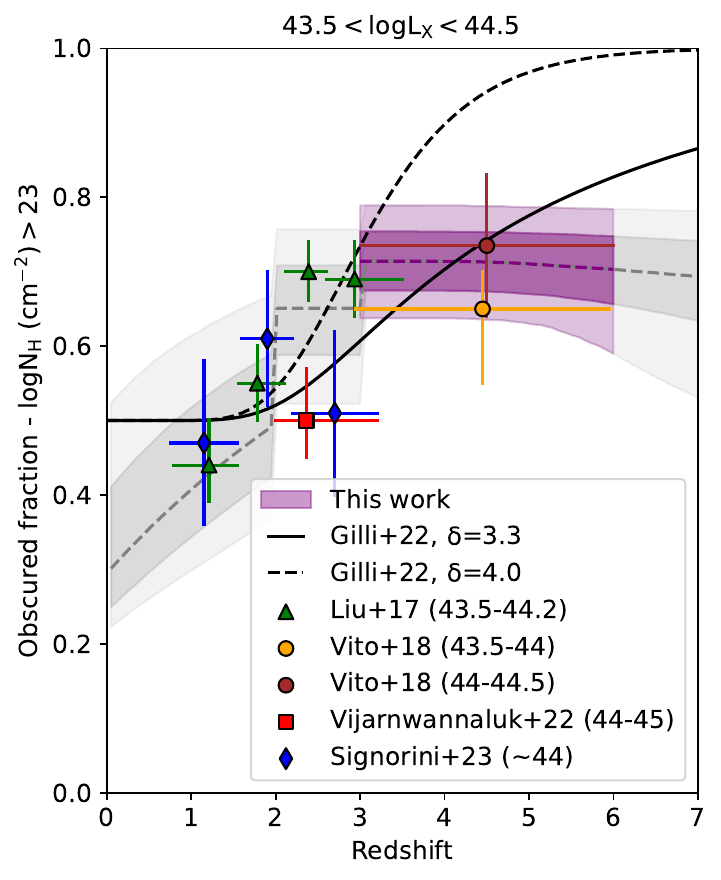} 
    \end{tabular}
\caption{Obscured fraction with $\log \mnh\geq 23$ versus redshift. Our results are shown with the purple shaded regions that represent the 68\% and 95\% confidence intervals of the best-fitting absorption function. In addition, we show the extrapolation (gray shaded regions) of the absorption function below and above the redshift range of our analysis. For reference, we show the results of \citet{Liu2017}, \citet{Vito2018}, \citet{klod2022} and \citet{Signorini2023} that span a wide range of redshifts. Inside the parenthesis, we provide the luminosity ranges used in the aforementioned studies. The lines are the predictions of the obscuration fraction originated on both ISM and torus components \citep{Gilli2022} for different values of the obscuration evolutionary factor, $\delta$.}
\label{gilli}
\end{figure}

In Fig.~\ref{gilli}, we present our results in addition to the obscured fractions given in previous studies in the literature \citep{Liu2017,Vito2018,klod2022,Signorini2023} at lower redshifts and having the same definition of the obscured fraction ($\rm logN_H (cm^{-2})>23$). We show only the AGN population with $43.5 \leq \log L_X \leq 44.5$ to compare properly with the aforementioned studies. \citet{Liu2017}, combining data from the 7 Ms CDF-S and the CCLS surveys, studied the obscuration properties of their sample and concluded that there is an evolution in the obscuration fraction from $z=1$ up to $z=3$. \citet{Vito2018} found that the obscured fraction defined as $N_{23-25}/N_{20-25}$ is almost constant at $\fabs=0.5$ in the redshift range within $z=3-5$ without any correction for the obscuration incompleteness. Applying the obscuration completeness, they found a value of $\fabs=0.65$ that is similar to our results. Concerning the AGN population with luminosities around $\log \LX\sim44$, they found an obscured fraction of $\sim 0.66$\% and $\sim 0.73$\% for the luminosity bins $43.5 \leq \log L_X \leq 44.0$ and $44.0 \leq \log L_X \leq 44.5$, respectively. \citet{klod2022}, using the deep multi-wavelength data of the XMM-SERVS catalogue inside the XMM-LSS region \citep{Chen2018}, suggested that 50\% of the sources are obscured at the cosmic noon ($z=2-3$). More recently, \citet{Signorini2023} studied the spectral properties of X-ray selected AGN in the J1030 Chandra field. They derived also the obscuration fraction of $\log L_X\sim44$ AGN in the redshift range $0.8\leq z \leq2.8$ with an average value of $\fabs=0.5-0.6$.

With the gray shaded regions we show the extrapolation of the absorption function below and above the redshift range of our analysis. For redshifts lower than $z=3$, as we discussed in Sect.~\ref{absfun}, we have adopted the formulations of \citet{Ueda2014} and \citet{klod2022}. Our predicted obscuration fraction is in agreement within the uncertainties to the previous works. In total, there is clear redshift evolution of the obscuration fraction of the AGN population at redshifts below the interval of our analysis.

This evolution can be driven by larger gas reservoirs observed in high-z AGN \citep{Carilli2013}. Indeed, \citet{Damato2020}, studying a sample of high-redshift AGN with $\log \mnh>23$, found that the contribution of the interstellar medium (ISM) of the host galaxies to the obscured fraction of the AGN is comparable to that estimated by X-ray or infrared estimations as in \citet{Circosta2019}. This may suggest that the host galaxy of the AGN can strongly affect the obscuration fraction, and especially, at high redshifts. The column density of the ISM at $z>3$ may be even 100 times larger than in the local Universe and could reach values close to the Compton-Thick regime at $z>6-8$. Such behavior has already been observed at $z>2$ where the ISM column density is consistent with the X-ray \nh\ values \citep{Gilli2022}. \citet{Gilli2022} modelled analytically at the same time both the large-scale (ISM) and small-scale (AGN torus) obscuration assuming clouds of different sizes, masses, and surface densities.

Comparing with different X-ray surveys, they concluded that the median ISM column density evolves with $N_{H,ISM}\propto (1 + z)^\delta$ where $\delta=3.3$. In addition, they found that the evolution of the characteristic cloud surface density with redshift ($\Sigma_{c,*}\propto (1 + z)^\gamma$), can be described with $\gamma=2$. \citet{Signorini2023} including their data and allowing for different values of the evolutionary factors found that $\gamma=2$ and $\delta=4$ best reproduce the data. In Fig.~\ref{gilli}, we show as well the estimations of the obscuration fraction predicted by the best \citet{Gilli2022} model with $\gamma=2$ and torus opening angle of $60\deg$. The latter was adopted since we are focused here on luminous AGN ($\log L_X \sim44$). The lines represent different values of the evolutionary factor $\delta$. Our results are consistent with an evolution of $\delta=3.3$ similarly to \citet{Gilli2022}. Hence, the large sample of high-z AGN used in our analysis is compatible with a scenario where the AGN obscuration fraction is dependent on the evolving ISM obscuration across cosmic time. However, we notice that the extrapolation of our obscured fraction towards higher redshifts ($z > \sim 6$) diverges from the ISM model by \citet{Gilli2022}. Larger samples of AGN, though, are needed at these redshifts (e.g., Sect.~\ref{conclusions}) to shed light on this context.

% %%%%%%%%%%%%%%%%%%%%%%%%%%%%%%%%%%%%%%%%%%%%%%%%%%%%%%%%%%%%%%%%%%%

\subsection{Black-hole accretion rate density}\label{section_bhad}

\begin{figure*}
\center
   % \begin{tabular}{c c c}
    \includegraphics[width=0.99\textwidth]{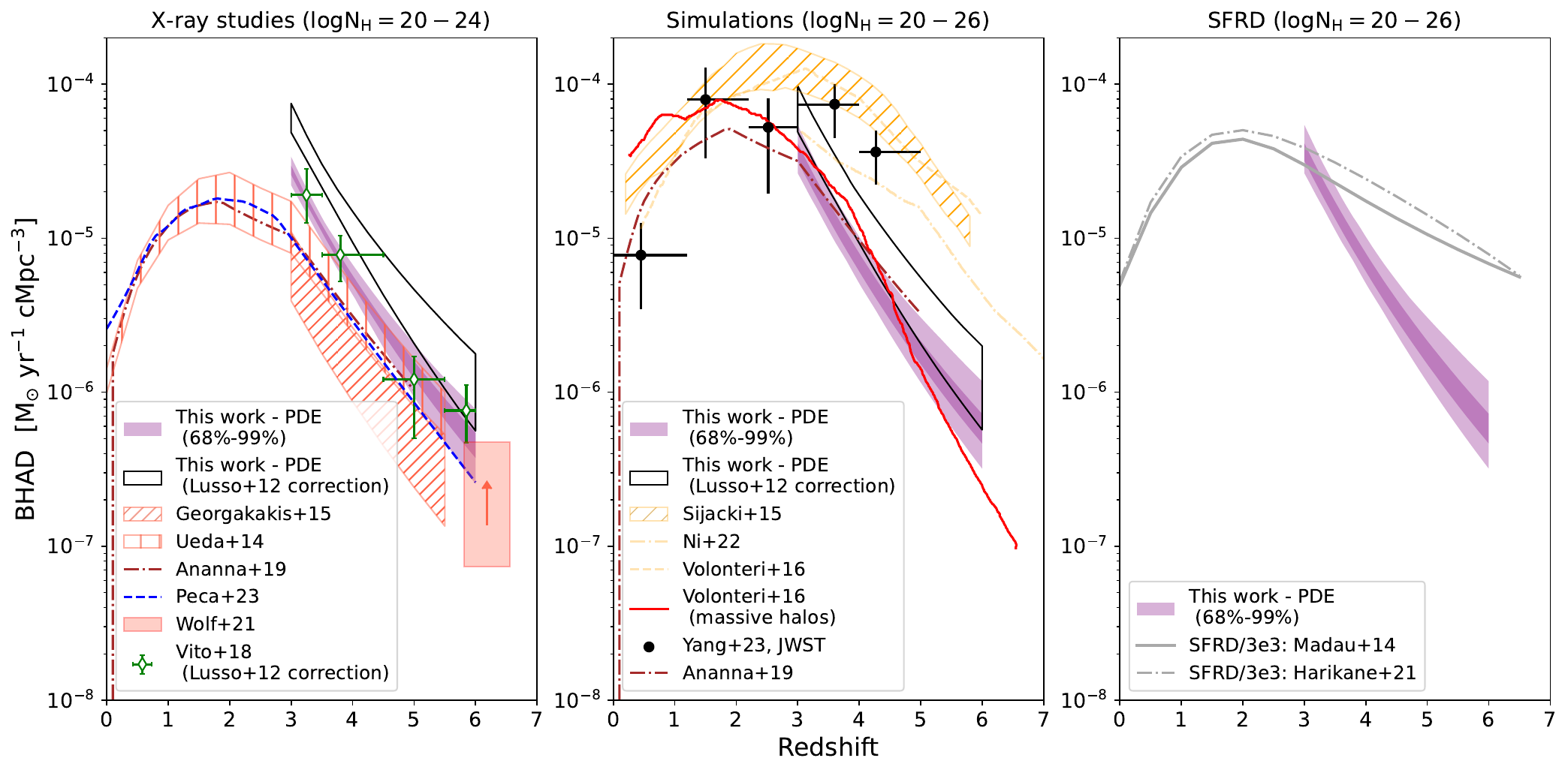} 
    % \includegraphics[width=0.31\textwidth]{plot_bhadxVOL_AUG2023-1.pdf} &
    % \includegraphics[width=0.31\textwidth]{plot_bhadxs_AUG2023-1.pdf} 
    % \end{tabular}
\caption{Redshift evolution of the black hole accretion rate density. The shaded regions represent the 68\% and 99\% confidence intervals of the BHAD using the best-fitting PDE model. From left to right, we compare our results with previous X-ray studies \citep{Georgakakis2015,Ueda2014,Vito2018,Wolf2021,Peca2023}, with simulations \citep{Volonteri2016,Ni2022,Sijacki2015} and with the star-formation rate density scaled down by a factor of 3000 \citep{Madau2014, Harikane2022}. The black regions in the first two panels indicate the 99\% confidence interval of the BHAD adopting the bolometric correction by \citet{Lusso2012} so we can compare directly to the results of \citet{Vito2018}. The black points in the middle panel correspond to the \textit{JWST} results by \citet{Yang2023}. The brown dashed-dotted line shows the predicted XLF derived by \citet{Ananna2019}.}
\label{bhad}
\end{figure*}

The black hole accretion rate density (BHAD) is fundamental to characterize effectively the growth of the AGN population. In this section, we derive the BHAD over cosmic time from our updated XLF and compare it to the predictions of theoretical models and the BHAD derived from several XLFs in the literature. Firstly, we converted the XLF into the bolometric luminosity function (BLF) using the luminosity-dependent bolometric correction by \citet{Duras2020}. In particular, we used the following equation:
\begin{equation}
\dv{\Phi}{\log \Lbol} = \dv{\Phi}{\log \LX} \times \dv{\log \LX}{\log \Lbol},
\end{equation}
where $\dv*{\Phi}{\log \Lbol}$ and $\dv*{\Phi}{\log \LX}$ are the bolometric and X-ray luminosity functions, respectively, and $\dv*{\log \LX}{\log \Lbol}$ is the derivative to convert the XLF to BLF. Then, we may compute the BHAD from the BLF:
\begin{equation}
\mathrm{BHAD} = \dfrac{1 - \epsilon}{\epsilon c^2} \times \int_{43}^{49} \Lbol \dv{\Phi}{\log \Lbol} \dd\log \Lbol,
\end{equation}
where $c$ is the speed of light and $\epsilon$ is the radiative efficiency to convert the energy into mass. In this work, we adopt $\epsilon =0.1$ similarly to previous works in the literature \citep{Hopkins2007}. The integral was calculated in the bolometric luminosity range $43 < \log \Lbol < 49$, corresponding to X-ray luminosities  of $42 \lesssim \log \LX \lesssim 47$ using the bolometric correction $(L_{\rm bol}/L_{\rm X})$ from \citet{Duras2020}.

In Fig.~\ref{bhad}, we show the BHAD based on our derived XLF as a function of redshift. The uncertainties were calculated using the output of the Bayesian analysis of the XLF. In addition, we have included the uncertainties of the \citet{Duras2020} correction parameters. We compare our results (left panel of Fig.~\ref{bhad}) for $\log N_{\rm H} =20-24$ with the BHAD computed from the XLF derived in previous X-ray studies \citep{Georgakakis2015,Ueda2014,Vito2018,Wolf2021,Peca2023} using the method described above. Since \citet{Vito2018} used the \citet{Lusso2012} bolometric correction, we show as well our BHAD adopting the latter correction (black region). Our BHAD is higher by a factor of about 2-4 compared to all the previous X-ray studies, while \citet{Wolf2021} derived an upper limit of the BHAD that agrees with ours at high redshifts. Furthermore, we compare our BHAD with the one predicted by \citet{Ananna2019}. As mentioned in the previous sections, in the $\log \mnh = 20-26$ bin, our results agrees within the uncertainties with their model. However, in the $\log \mnh = 20-26$ we find a much higher value. This difference can be ascribed to the fact that in their analysis the predicted Compton-Thick fraction is $\sim$50\% that is much higher compared to our value ($F_{CTK}=0.17_{+0.07}^{-0.09}$).

In the middle panel of Fig.~\ref{bhad}, we compare our derived BHAD with the ones predicted by various large-scale cosmological simulations \citep{Volonteri2016,Ni2022,Sijacki2015}. When we consider the full AGN population predicted in simulations, we find more or less one order of magnitude difference. However, as discussed in Sect.~\ref{space}, the X-ray sample consists of systems with stellar masses of $\rm \geq 10^{10}~M_\odot$. Restricting the simulated AGN only to massive systems, this difference should be decreased. Indeed, using the predictions of \citet{Volonteri2016} for massive dark matter halos ($\rm \geq 5\times 10^{11}~M_\odot$) that roughly correspond to a cut in the stellar mass of the host galaxies at $\rm 3\times 10^{10}~M_\odot$ \citep{Dubois2015}, the BHAD coming from X-rays is in a very good agreement with the simulations. So, in order to compare properly the BHAD derived from X-ray studies and the simulations, one should be very conscious of the selection biases (e.g., stellar mass, obscuration, etc.).

One explanation for this difference in BHAD is that simulations may produce too much accretion in small systems. On the other hand, \citet{Yang2023}, using mid-infrared observations from the Mid-Infrared Instrument (MIRI) onboard \textit{JWST}, found that the BHAD is about 0.5 dex higher than that of previous X-ray studies, while it is more consistent with the simulations. If we compare it with our results at $z \geq 3$, we find a difference of about 0.8 dex and 0.5 dex when using the \citet{Duras2020} and \citet{Lusso2012} bolometric correction, respectively. They ascribe this difference to the fact that X-rays may miss a large population of heavily obscured AGN and hence the simulations over-predict the number counts. If this is the case, it means that even our correction of the obscuration incompleteness is not sufficient enough.

However, in \citet{Maiolino2023} and \citet{Lyu2023} the vast majority of the \textit{JWST} IR selected AGN in the redshift range $3.0 \leq z \leq 6.0$ are hosted by low-mass galaxies ($\rm < 10^{10}~M_\odot$). The latter study also suggests that the low-mass galaxy population is comparable to that of the high-mass galaxies at least in the cosmic noon. We argue here that the X-ray population may miss a small fraction of heavily obscured AGN compared to the SED selected AGN, though the large discrepancies in BHAD arise from the different AGN populations selected in each selection method, and especially, different host-galaxy properties. A straight-forward test would be to directly compare the host-galaxy properties between our sample and the \textit{JWST} high-z AGN \citep{Yang2023}. Since the stellar mass estimates of the \textit{JWST} sources are not available, we examine this issue by using the relation between the stellar mass of the host and the black-hole accretion rate ($BHAR$). \citet{Yang2017,Yang2018} have shown that $BHAR$ strongly depends on the stellar mass of the host galaxy. Hence, we compare the $BHAR$ of the high-z \textit{JWST} sources in \citet{Yang2023} with the SED fitting results of the X-ray detected sources in \citet{Pouliasis2022b}. For the latter, we used similar approach to the former. We find that the X-ray detected sources have $\log BHAR = -1,+3$, while the mid-IR AGN ranges between $\log BHAR = -3,-0.5$. This supports further our argument that X-ray studies are focused on identifying AGN hosted by larger-mass galaxies, while the mid-IR observations unveil in a complementary manner the low-mass regime. Furthermore, by comparing the bolometric luminosities, we find significant differences between the selection methods. In particular, using the \citet{Duras2020} bolometric correction, we derived a median value of $\log \Lbol = 45.9 \pm 0.75$. \citet{Yang2023}, using the disk luminosity as a proxy for the angle-averaged bolometric luminosity, found $\log \Lbol = \sim 44.5$ in the redshift range $z=3-8$ (their Fig.~6). In addition, \citet{Lyu2023} derived a similar value of $\log \Lbol = \sim 44.6$ in the redshift range $z=3-4$.

Finally, we compare the BHAD evolution with the star-formation rate density (SFRD) evolution. Above redshift two there is a decline at both BHAD and SFRD. Many simulations \citep{Habouzit2021,Zhang2023} have shown that at higher redshifts ($z>3$), the ratio BHAD/SFRD drops, indicating that the AGN growth evolves more rapidly than the galaxies. On the right panel of Fig.~\ref{bhad}, we plot our derived BHAD with the SFRD predicted by \citet{Madau2014} and \citet{Harikane2022} scaled down by a factor of 3000 to match approximately our BHAD at $z=3$. The shape of the SFRD from the aforementioned studies is qualitatively consistent with those of the simulations. We found that the BHAD/SFRD ratio drops (as in simulations) from $4\times10^{-4}$ at $z=3$ to $7\times10^{-7}$ at $z=6$. These results are in contrast to the findings of \citet{Yang2023}, where they found that the BHAD/SFRD ratio increases at $z \geq 3$.

%--------------------------------------------------------------------

%-----------------------------------------------------------------

\section{Summary and conclusions}
\label{conclusions}
We have built the largest sample ($>600$) of X-ray selected AGN at high redshifts ($z>3$). The sample is compiled using fields with different areas and sensitivity depths (CDF-S,  CDF-N, CCLS, XMM-XXL north). About one third of our sources have spectroscopic redshifts available. In this paper, we place tight constraints on the X-ray luminosity and the absorption functions. The advantage of this work is that we derived the X-ray spectral properties (luminosities, absorbing column densities) using a Bayesian technique in a consistent way for all sources taking into account the photometric redshift uncertainties. Our main results can be summarised as follows.
\begin{itemize}

\item The XLF is described by a double power-law which evolves according to a pure density evolution model similar to what is witnessed in optical wavelengths. However, a luminosity dependent density evolution model cannot be securely ruled out.

\item The vast majority of our sources are heavily obscured with column densities $\log \mnh > 23$. Our results confirm previous findings from \citep{Vito2018} in the \Chandra\ Deep Fields. We find no luminsity dependence on the obscured fration, while by combining our results with those at lower redshifts, we find an evolution of the obscuration with redshift. This could be explained by ISM obscuration in addition to that of the torus \citep[e.g.][]{Gilli2022}.

\item The BHAD derived through our analysis is higher compared to previous X-ray studies, while it is roughly in agreement with the simulations if one uses a cut in the stellar mass of the host galaxies. Comparing with the BHAD derived using \textit{JWST} data, we conclude that IR-selected AGN concern sources with lower bolometric luminosities hosted by galaxies with $\rm \leq 10^{10}~M_\odot$, while X-rays probe AGN hosted by larger systems.

\end{itemize}

Currently the eROSITA all-sky survey is detecting a few million AGN. At the same time, \XMM\ has detected roughly one million serendipitous sources covering $\sim1300\deg^2$ \citep{Webb2023}, while the latest Chandra Source Catalog Version 2.1 (CSC 2.1) will include about 390 thousands unique X-ray sources detected by the Chandra X-ray observatory \citep{Evans2010,Evans2020}. All of the above surveys are a treasure trove for studies of high redshift AGN. These surveys are highly complementary. eROSITA owing to the large area covered is expected to detect the most luminous AGN occupying the bright end of the luminosity function. On the other hand \XMM\ and Chandra will explore luminosities near and below the break of the luminosity function, respectively. Because of the combination of their energy band-passes and the flux depth probed, eROSITA will detect primarily unobscured AGN while \XMM\ and Chandra will probe more efficiently the obscured AGN population at high redshift.

Evidently, because of the faintness of the optical counterparts of the high redshift AGN, excellent quality optical data are necessary in order to fully exploit the high redshift content of these surveys. To have a deeper knowledge on the physical properties governing the growth and evolution of super-massive black holes over cosmic time, large AGN samples are needed to map those populations missed by the current X-ray missions (low-luminosity, obscured sources at medium-high redhsifts). In the future, ESA's New ATHENA \citep[advanced telescope for high-energy astrophysics,][]{Nandra2013} X-ray mission will revolutionise the studies of high redshift AGN owing to its unprecedented sensitivity. In addition, the next-generation X-ray telescopes, such as the Advanced X-ray Imaging Satellite \citep[AXIS,][]{Mushotzky2019,Marchesi2020}, the Survey and Time-domain Astrophysical Research eXplorer (STAR-X\footnote{\url{http://star-x.xraydeep.org/}}), a Medium Explorer mission selected by NASA, and the High Energy X-ray Probe \citep[HEX-P,][]{Madsen2018,Madsen2023} will offer new opportunities to detect and study obscured sources at high redshifts.

\begin{acknowledgements}
     The authors are grateful to the anonymous referee for a careful reading and helpful feedback. EP, AR and IG acknowledge financial support by the European Union's Horizon 2020 programme "XMM2ATHENA" under grant agreement No 101004168. The research leading to these results has received funding (EP and IG) from the European Union's Horizon 2020 Programme under the AHEAD2020 project (grant agreement n. 871158).  
     
     The XXL survey\footnote{\url{http://irfu.cea.fr/xxl}} is an international project based around an \XMM\ Very Large Programme surveying two $25\deg^2$ extragalactic fields .
     Multi-band information and spectroscopic follow-up of the X-ray sources are obtained through a number of programmes, \url{http://xxlmultiwave.pbworks.com}. This research made use of Astropy, a community-developed core Python package for Astronomy \citep[\url{http://www.astropy.org},][]{astropy2018}. This publication made use of TOPCAT \citep{Taylor2005} for table manipulations. The plots in this publication were produced using Matplotlib, a Python library for publication quality graphics \citep{Hunter2007}.
     
     Based on observations obtained with XMM-Newton, an ESA science mission with instruments and contributions directly funded by ESA member states and NASA. 
     
     This research has made use of data obtained from the \Chandra\ Data Archive and the \Chandra\ Source Catalog, and software provided by the \Chandra\ X-ray Center (CXC) in the application packages CIAO and Sherpa.

    The Hyper Suprime-Cam (HSC) collaboration includes the astronomical communities of Japan and Taiwan, and Princeton University. The HSC instrumentation and software were developed by the National Astronomical Observatory of Japan (NAOJ), the Kavli Institute for the Physics and Mathematics of the Universe (Kavli IPMU), the University of Tokyo, the High Energy Accelerator Research Organization (KEK), the Academia Sinica Institute for Astronomy and Astrophysics in Taiwan (ASIAA), and Princeton University. Funding was contributed by the FIRST program from the Japanese Cabinet Office, the Ministry of Education, Culture, Sports, Science and Technology (MEXT), the Japan Society for the Promotion of Science (JSPS), Japan Science and Technology Agency (JST), the Toray Science Foundation, NAOJ, Kavli IPMU, KEK, ASIAA, and Princeton University. This paper makes use of software developed for the Large Synoptic Survey Telescope. We thank the LSST Project for making their code available as free software at  http://dm.lsst.org. This paper is based [in part] on data collected at the Subaru Telescope and retrieved from the HSC data archive system, which is operated by Subaru Telescope and Astronomy Data Center (ADC) at National Astronomical Observatory of Japan. Data analysis was in part carried out with the cooperation of Center for Computational Astrophysics (CfCA), National Astronomical Observatory of Japan.

\end{acknowledgements}



% \nocite{*}
\bibliographystyle{aa}
%\bibliography{bibektoras}
\bibliography{aanda} % your references Yourfile.bib

\begin{appendix}

%%%%%%%%%%%%%%%%%%%%%%%%%%%%%%%%%%%%%%%%%%%%%%%%%%%%%%%%%%%%%%%%%%%%%%%%%%%%%%%%%%%

\section{XMM-XXL-N redshift estimations}\label{lephare}

In the XMM-XXL northern field, we used an internal release obtained with the V4.2 XXL pipeline that has been used in \citet{Pouliasis2022a,Pouliasis2022b}. This version is expected to be superseded by the final catalogue, V4.3. The initial catalogue contains 15547 X-ray sources. Restricting our sample to point-like sources only detected in the soft band, we ended up with 13610 X-ray sources. Since our main goal was to use the HSC data to increase the completeness of the identifications and the photometric redshift completeness, we restricted the X-ray sample into the HSC coverage. Thus, the area of the XMM-XXL-N field that overlaps with the HSC data (after excluding the HSC masked areas) is about $21\deg^2$. In this area, there are 10232/13610 ($\sim$75\%) X-ray sources detected in the soft band. In the next sections, we present the methodology and the data we used to determine the redshift of the sources.

\subsection{Spectroscopic redshifts}
The XMM-XXL-N field has been covered by several spectroscopic surveys targeting extra-galactic sources that include both AGN and galaxy samples. The majority of the targets are pre-selected in the optical or UV wavelengths, but there are many surveys dedicated only to X-ray selected sources. In our analysis, we used the spectroscopic information gathered by the HSC team and was already associated with the HSC photometric catalogue. This catalogue contains spectroscopic redshifts from the PRIsm MUlti-object Survey \citep[PRIMUS][]{Coil2011,Cool2013} in the sub-region XMM-LSS ($\sim2.88\deg^2$) of the XXL-N, the Galaxy And Mass Assembly \citep[GAMA,][]{Liske2015}, the VIMOS VLT Deep Survey \citep[VVDS,][]{Lefevre2013}, the VIMOS Public Extragalactic Survey \citep[VIPERS,][]{Garilli2014} and the latest data release of \citep[SDSS-DR16,][]{Ahumada2020}. SDSS-DR16 that is the fourth release of the Sloan Digital Sky Survey IV and includes the previous data releases DR12 and DR14 \citep{Alam2015,Paris2018}. Additionally, we made use of the spectroscopic catalogues of X-ray detected sources derived in \citet{Menzel2016} and \citet{Akiyama2015}. For the latter, the spectroscopic catalogues were matched to the optical positions in our sample with a radius of 1 $\arcsec$. Initially, we selected all spectroscopic redshifts that have secure measurements according to the flags provided in each catalogue. Then, we included in the spectroscopic sample sources with less reliable spectroscopic redshift estimations but that do agree with the photometric redshifts (next section) within $|\Delta z|/(1+z_{\rm spec})>0.10$ following \citet{Marchesi2016optical} methodology. We ended up with 3673 sources with available spectroscopic information (1414 point-like and 2259 extended sources) with 70 of them being at $z>3$.

\subsection{Photometric redshifts}

\begin{figure}
\center
   \begin{tabular}{c}
    \includegraphics[width=0.47\textwidth]{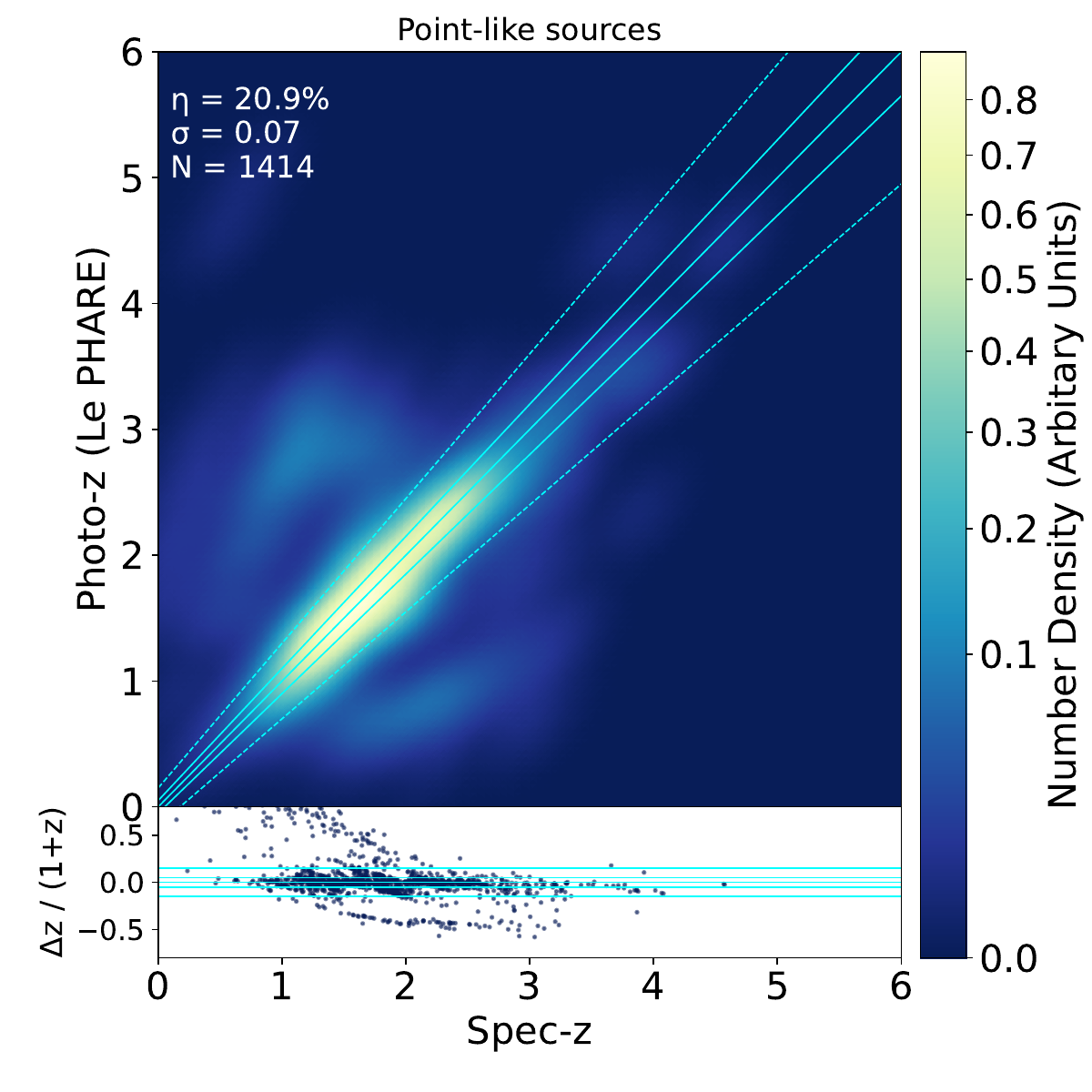} \\
    \includegraphics[width=0.47\textwidth]{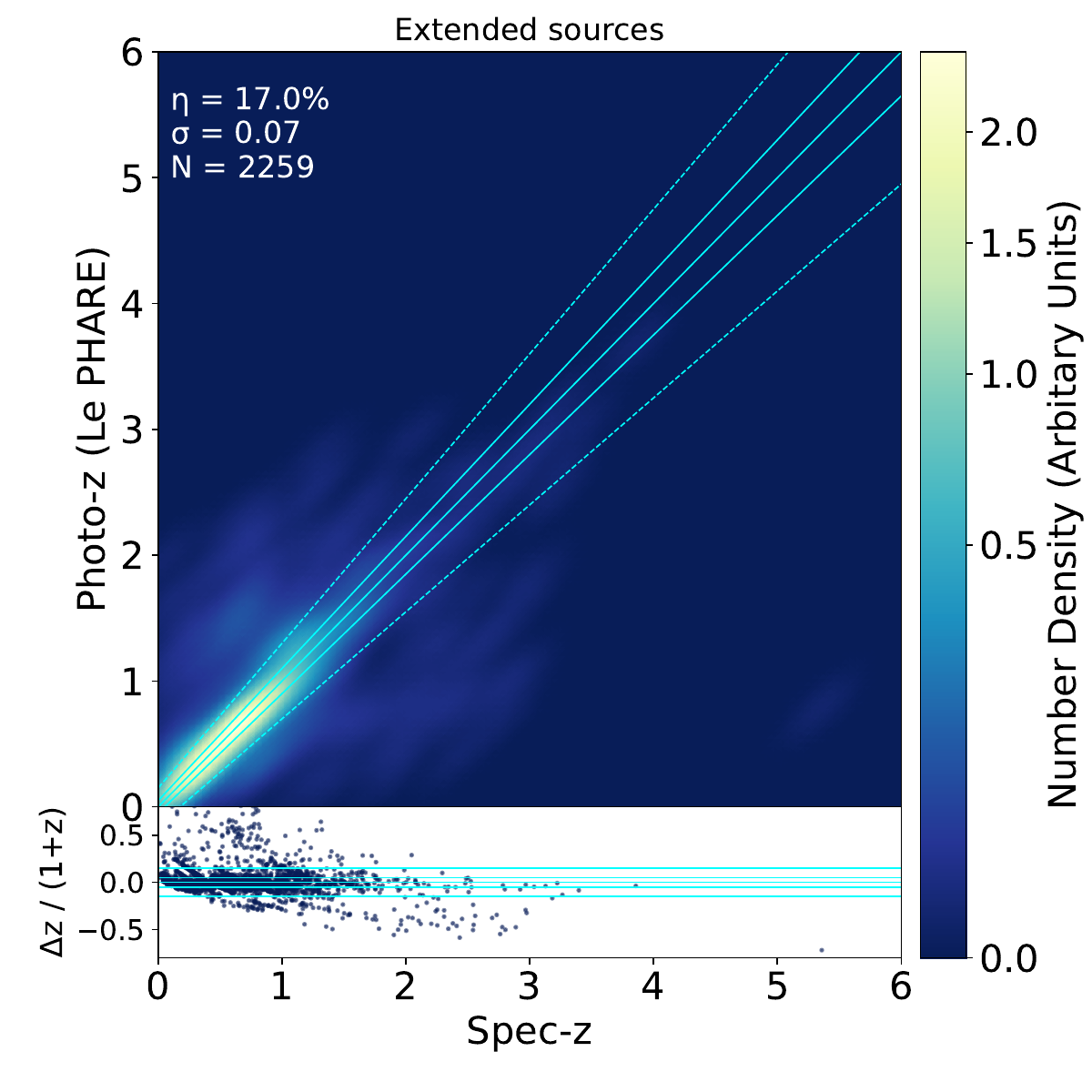} 
    \end{tabular}
\caption{Photometric versus spectroscopic redshifts for the X-ray sources that have available spec-z information. The dotted lines represents the limits of the catastrophic outliers. }
\label{photoz_accuracy}
\end{figure}

For the X-ray sources that do not have available spectroscopic information, we derived the photometric redshifts using the LePHARE SED fitting algorithm \citep{Arnouts1999,Ilbert2006}. LePHARE is able to provide the best fit among different templates using a $\chi^2$ minimization procedure between the observed and the model photometry of the SEDs. As proposed in many previous studies \citep[e.g.,][]{Salvato2022,Ilbert2009}, it is important to run LePHARE with different templates depending on the morphological type of a source (point-like or extended) to avoid degeneracies between the parameters of the models. Hence, for point-like sources we used a library of templates similar to that applied by \citet{Salvato2022}, while for the X-ray sources that have an extended morphology in the optical images, we used a library of templates as in \citet{Salvato2009,Salvato2011}. In addition, we applied different luminosity priors for these samples. In particular, following previous studies \citep[e.g.,][and references therein]{Ananna2017}, we used an absolute magnitude of $-24<M_{HSC_g}<-8$ and $-30<M_{HSC_g}<-20$ for the extended and point-like sources, respectively.

To construct the observed SEDs of the X-ray sources, we used the photometry described in \citet{Pouliasis2022a}. Furthermore, we complemented this data-set with deeper observations in the optical, near-infrared and mid-infrared wavelengths. In particular, we added the deep layer of the optical HSC-SSP PDR2 catalogue \citep{Aihara2019} and the latest data releases of the near-infrared VISTA Hemisphere Survey \citep[DR5,][]{Mcmahon2013}, VIKING \citep[DR5,][]{Edge2013}, VISTA Deep Extragalactic Observations (VIDEO) Survey \citep[][]{Jarvis2013} and the DXS and UDS surveys of the UKIRT Infrared Deep Sky Survey \citep[DR11,][]{Lawrence2007}. In addition, we have included in our analysis deeper mid-infrared data from the latest data releases of the Spitzer SWIRE \citep{Lonsdale2003} and the SERVS \citep{Mauduit2012} surveys.

To have an estimation of the photometric redshift accuracy, we used the sub-sample of the X-ray sources with available spectroscopic information. In particular, we used the traditional statistical indicators: the normalised median absolute deviation $\sigma_{\rm NMAD}$ \citep{Hoaglin1983,Salvato2009,Ruiz2018} and the percentage of the catastrophic outliers $\eta$ \citep{Ilbert2006,Laigle2016} that can be defined as follows:
\begin{equation}
\sigma_{\rm NMAD}=1.4826\times median \left( \frac{|\Delta z-median(\Delta z)|}{1+z_{\rm spec}}\right)~
\end{equation}
and
\begin{equation}
\eta (\%)=\frac{N_{\rm outliers}}{N_{\rm total}}\times100,
\end{equation}
where $\Delta z=z_{\rm phot}-z_{\rm spec}$, $N_{\rm total}$ is the total number of sources and $N_{\rm outliers}$ is the number of the outliers. An object is defined as an outlier if it has $|\Delta z|/(1+z_{\rm spec})>0.15$. Figure~\ref{photoz_accuracy} shows the comparison between spectroscopic and photometric redshift estimations for the samples of point-like (upper) and extended (lower) sources.

We obtained $\eta$=20.9\% and $\sigma_{\rm NMAD}$=0.07 for the point-like sources and $\eta$=17\% and$\sigma_{\rm NMAD}$=0.07 for the extended sources. Breaking down the sample, we find that the point-like sources have a higher accuracy at higher redshifts ($z>2$) compared to the extended sources, and, vice versa. Also, the LePHARE algorithm performs equally well at all magnitudes for the extended sample, but regarding the point-like sources there is an increase of the fraction of outliers and the scatter at the very faint magnitudes ($HSC_i\geq24$). However, at these magnitudes the number of sources is very low compared to the bright ones. Furthermore, when considering only the spec-z sample at high redshift ($z\geqslant3$), the accuracy significantly improves with $\eta$=19\%. This fraction decreases to $\eta=\sim 10$\% for sources with $z>3.5$ up to zero for sources with $z>4$. This is an improvement compared to previous works, such as \citet{Salvato2022} where they found $\eta$=27.3\% for their $z\geqslant3$ sample.

\section{Goodness of fit, informational gain and reliability of X-ray spectral fits}
\label{app_xfit}

In this appendix, we present in detail the methodology we followed to evaluate the goodness of our X-ray spectral fit analysis (App.~\ref{xfit_gof}). We also explore the informational gain we obtained for the model parameters (App.~\ref{xfit_dkl}) and the reliability of our results (App.~\ref{xfit_rel}), particularly in the case of low-count spectra.

\begin{table}[t]
\caption{Input parameters for the spectral model used for the reliability simulations.}
\label{tab:sims_params} 
\centering
\begin{tabular}{l c}
\hline\hline
 Parameter & Value \\
\hline
 \multicolumn{2}{c}{Galactic absorption} \\
 Galactic \nh & $3\times10^{22}~\mathrm{cm^{-2}}$ \\
\hline
 \multicolumn{2}{c}{UXClumpy} \\
\hline
 $\log \mnh$ & 21.5, 23.5, 24.5 \\
 phoindex & 1.95 \\
 ecut & 300~keV \\
 torsigma & 28~deg \\
 ctkcover & 0.4 \\
 theta\_inc & 18.5~deg \\
 redshift & 3.5 \\
 normalization\tablefootmark{a} & $6.174\times10^{-5}$ \\
 scattering fraction & 0.01 \\
\hline
\hline                                             
\end{tabular}
\tablefoot{\tablefoottext{a}{Normalization value selected such as $\log \LX = 45$.}}
\end{table}

\begin{figure*}
\centering
    \includegraphics[width=\linewidth]{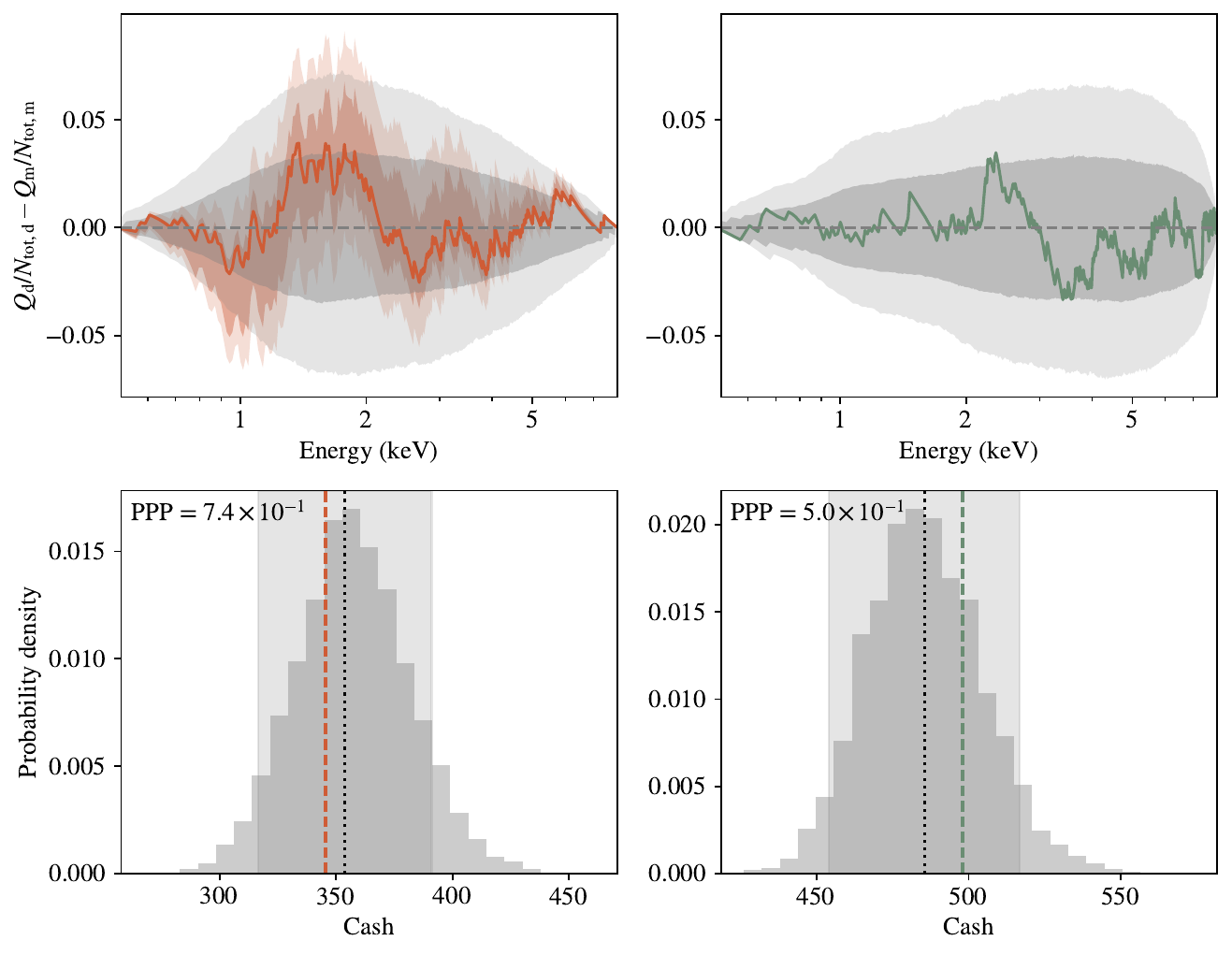}
\caption{Example of posterior predictive checks for the \Chandra~source CCLS LID 460. Panels on the left column show the results for the source data, while the right column correspond to the background data. \textbf{Top panels:} Differential QQ-plots. The gray shaded areas show the one-sigma and two-sigma percentiles for 5000 X-ray spectral simulations using a random sampling of the posterior distribution. The solid lines correspond to the best-fit model obtained using BXA, and the red shaded areas are the one-sigma and two-sigma percentiles from the posterior distribution. \textbf{Bottom panels:} The gray histograms show the distribution of the Cash statistic for the simulations. The vertical, dashed lines show the Cash value for our best-fit model. The vertical, dotted lines and the vertical shaded areas show, respectively, the expected Cash value ($C_e$) for the best fit-model and the  corresponding 90 per cent confidence region estimated using the \citet{Kaastra2017} method. The values in the upper-left corner of each panel are the posterior predictive p-values estimated using the simulations.}
\label{fig:ppcheck}
\end{figure*}

\begin{figure*}%
\begin{minipage}{0.82\linewidth}%
    \includegraphics[width=\textwidth]{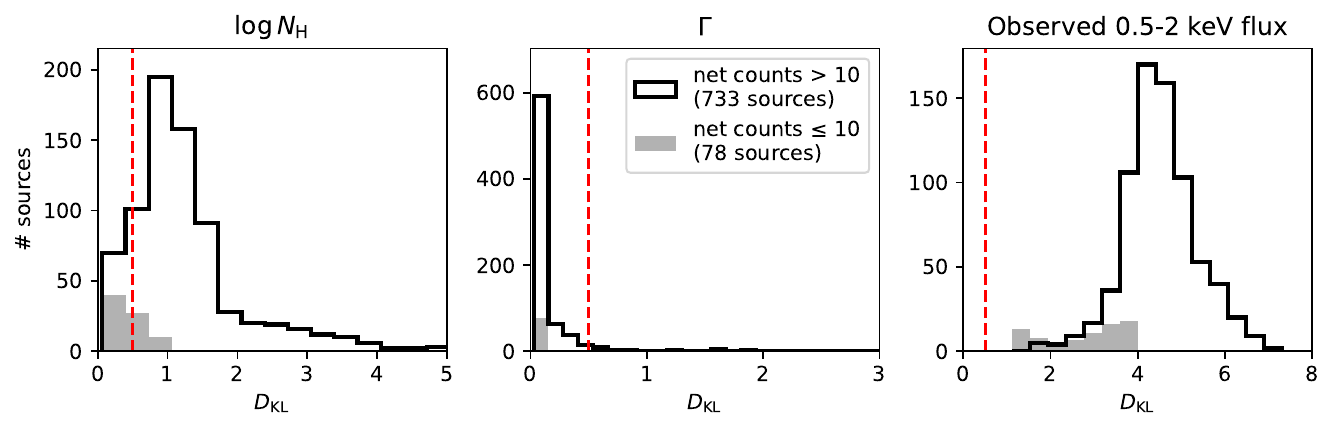}\\
    \includegraphics[width=\textwidth]{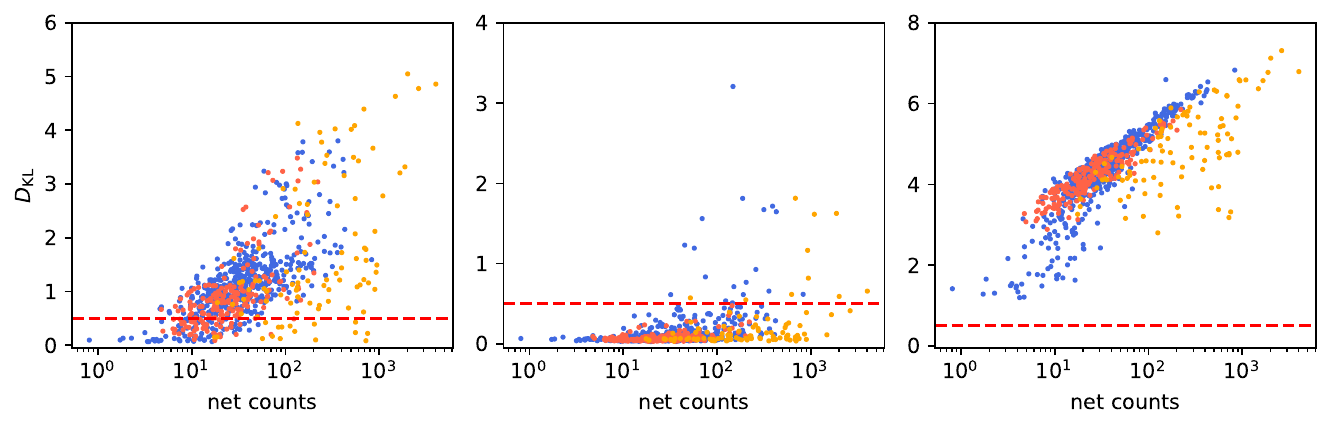} 
\end{minipage}%
%\qquad
\begin{minipage}{0.15\linewidth}%
    \includegraphics[width=1.25\textwidth]{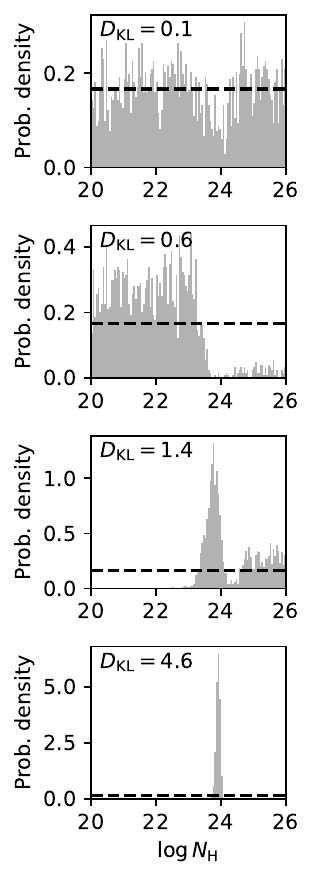}
\end{minipage}%
\caption{\textbf{Top row:} Distribution of the Kullback–Leibler divergence for Hydrogen column density, photon index and observed X-ray flux. The solid, black lines show the distribution for sources with more than ten net counts in their X-ray spectra, Gray shaded areas show the distribution for sources with $\leq10$ net counts. Vertical, dashed, red lines show $D_\mathrm{KL} = 0.5$.
\textbf{Bottom row:} The panels show $D_\mathrm{KL}$ versus net counts for the same three parameters in the top row. Symbols are color-coded according to the source survey (blue: XXL-North; red: CCLS; orange: CDF). The horizontal, dashed, red lines again show $D_\mathrm{KL} = 0.5$. \textbf{Right column:} Four examples of the $\log\mnh$ posterior distribution (gray histograms) obtained from our X-ray spectral fits, for different $D_\mathrm{KL}$ values. The horizontal, dashed lines show the flat prior distribution we assumed in our analysis.}%
\label{fig:dkl}%
\end{figure*}

\begin{figure}
\centering
    \includegraphics[width=\linewidth]{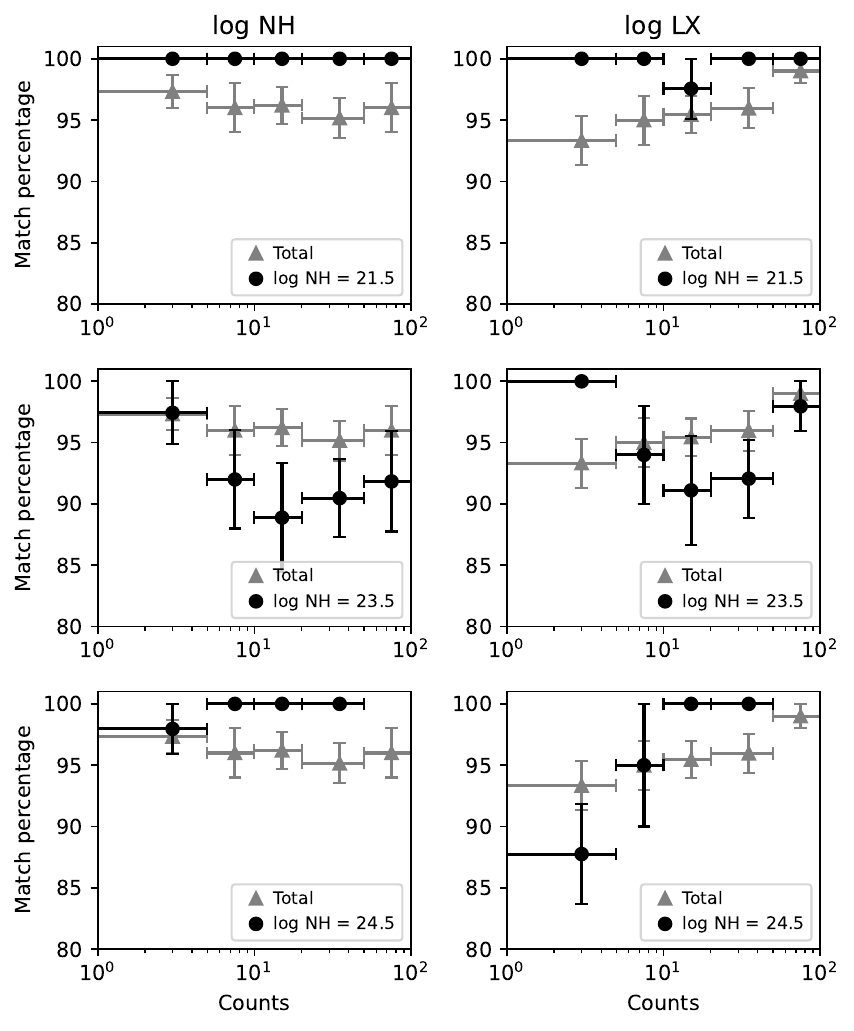}
\caption{Dependence of the match percentage with the number of counts in the X-ray spectra for the Hydrogen column density (left column) and the intrinsic 2-10~keV luminosity (right column). The top/middle/bottom rows correspond to the results for sources with $\log \mnh=21.5/23.5/24.5$, respectively (black circles). In all panels we have also included the results for the total sample, without splitting in $\log \mnh$~(gray triangles). Error bars correspond to one-sigma uncertainties, calculated through bootstrapping.}
\label{fig:reliability}
\end{figure}

\subsection{Goodness of fit}
\label{xfit_gof}

The goodness of fit (GoF) evaluates how well the selected model in a fitting procedure reproduces the observed data. While no statistical test is able to confirm that the model is correct, if the GoF does not reach a certain significance level, the model can be rejected. The Cash statistic we used for our BXA X-ray spectral fits does not provide a GoF test (unlike e.g., the $\chi^2$ statistics). However, \citet{Kaastra2017} proposed a Cash-based test for model testing: for a given model, it is possible to calculate the expected Cash value ($C_e$) and its variance ($C_v$). For a selected significance level, it is then possible to define an interval centered around $C_e$. If the Cash value for the best-fit model is outside this interval, then the model is rejected. We performed this test for the X-ray spectral fits for all sources in our sample. Using a 90 per cent significance level, the model was not rejected for all sources.

Alternatively, a robust, fully Bayesian approach for estimating the GoF is through posterior predictive checks \citep[see e.g., Chapter 6 of][]{Gelman2014}. The basic idea of the method is to do a random sampling of the posterior distribution obtained during the fitting procedure, and then create simulated data for this random sampling. This simulated data set can be compared to the actual observed data set via visual or statistical methods. For example a qualitative, visual test can be done via QQ-plots \citep{Buchner2014,Buchner2023}, where the cumulative count distributions of the data and the model are compared. In the top panels of Fig.~\ref{fig:ppcheck}, we show the differential QQ-plots using the BXA results for a \Chandra~source selected from our sample. The plots show the difference for the normalized cumulative distribution of counts between the data and the model, as a function of the observed energy. The gray, shaded areas show the one- and two-sigma percentiles for 5000 simulated spectra using a sampling of the UXClumpy model parameters from the BXA posterior distribution, and the corresponding instrumental setup for the observed spectrum (same response matrices and exposure times) and background model. The red, shaded areas in the left panel correspond to the results of the actual posterior distribution. If the results for the observed data show strong deviations from the area defined by the simulations in the QQ-plots, then the model should be rejected.

A quantitative check can be done using posterior predictive p-values \citep{Meng1994,Gelman2014}. For each simulation in the posterior sampling, the corresponding Cash statistic is calculated. A p-value can be calculated by comparing the Cash distribution of the simulations with the distribution for the actual, observed data. If the obtained p-value is below a pre-selected significance level, then the model should be rejected. The lower panels of Fig.~\ref{fig:ppcheck} show the Cash distributions for the 5000 simulations (gray histograms) and the Cash statistics for the best-fit model (vertical dashed lines). The numbers in the upper-left corner of each panel correspond to the two-sided posterior predictive p-value we estimated. The vertical gray shaded areas show the 90 percent confidence intervals predicted by \citet{Kaastra2017}, and the black, dotted vertical lines are the corresponding expected Cash values.

We compared the results of the \citet{Kaastra2017} method and the posterior predictive checks for a subset of sources, including low and high count spectra, and in all cases we found consistent results between both methods. Posterior predictive checks are computationally very expensive. Given the size of our sample we decided to use only the \citet{Kaastra2017} method for the whole sample.

\subsection{Informational gain}
\label{xfit_dkl}
In this section, we investigate the knowledge we obtained for different X-ray spectral model parameters through our BXA spectral fits. This is in particular interesting in the case of low-count spectra, where in principle the model parameters are not tightly constrained. To this end, we have used the Kullback–Leibler divergence \citep[$D_\mathrm{KL}$,][]{KLdiv1951}, which is a statistic that quantifies the difference between the posterior and prior distributions for a given parameter (see Sect.~\ref{fit}). We followed the method presented in \citet{Buchner2022} for the numerical calculation of $D_\mathrm{KL}$. As a rough, intuitive way of understanding the informational gain obtained for a given parameter, if $D_\mathrm{KL}=0.5, 1, 2$, then the shrinking factors between a Gaussian prior and a Gaussian posterior are, respectively, $\sim2$, $\sim3$ and $\sim6.5$ (see Fig.~1 of \citealt{Buchner2022}).

We calculated the $D_\mathrm{KL}$ for three parameters: $\log \mnh$, the photon index $\Gamma$, and the observed 0.5-2~keV flux. While the latter is not a direct parameter of the UXClumpy model, it is a derived quantity directly linked with the overall normalization of the model. We compared the derived posterior distribution of the X-ray flux with a flat prior covering the full range of observed fluxes in our sample. 

Figure~\ref{fig:dkl} show our estimates of the informational gain we obtained for our X-ray fitting results. The panels in the top row show the $D_\mathrm{KL}$ distribution for the three parameters of interest, divided between sources with equal or less than ten net counts (gray histograms), and more than ten counts (solid, black lines). The lower panels show the $D_\mathrm{KL}$ estimates versus the net spectral counts for each source in our sample. 

The plots show how for example the informational gain for the X-ray fluxes is high ($D_\mathrm{KL} > 1$) for all sources, including those with low-count spectra. This is expected, since even a detection barely above the sensitivity limit of the survey is enough to put a strong constrain in the flux of the source. We can also see how the informational gain for $\Gamma$ is very limited, with most sources showing $D_\mathrm{KL} \leq 0.5$, and it shows no strong dependence with the number of counts. There are two major reasons for this result: on one hand, the prior we selected for the photon index correspond to the $\Gamma$ distribution for the overall population of AGNs; on the other hand, the degeneracy between \nh~and $\Gamma$ adds a dispersion level to the posterior distribution of $\Gamma$ that is difficult to reduce given the redshifts of our sources and the observed energy range of the spectral data.

In the case of the Hydrogen column density, our results show that for most of the sources with spectral counts above 10, the informational gain is significant, with $D_\mathrm{KL} > 0.5$. For sources with very low counts ($< 5$) there is no gain, the posterior and prior distributions are very similar. It is nevertheless interesting that for sources between about five and ten counts some of them show a significant informational gain, demonstrating than in such cases the posterior distribution can be constrained to a certain level. In the next section we will explore how reliable are these constrains for $\log \mnh$ in the case of low-count spectra.

\subsection{Reliability}
\label{xfit_rel}
In this section, we explored how the number of counts available in the X-ray spectra affects the reliability of the fitting results, particularly in the low counts regime. To this end, we have followed a methodology similar to the one presented in \citet{Peca2023}. Using spectral simulations we evaluated the performance of the BXA fitting results. We simulated \XMM~EPIC-PN spectra\footnote{We used the response and ancillary matrices provided by SIXTE, an X-ray observation simulation software \citep{sixte}.} assuming the  UXClumpy model we used for our spectral analysis. We used three different values for the absorption level, with $\log \mnh = 21.5, 23.5, 24.5$, and five exposure time values: 2, 5, 10, 20 and 50~ks. The remaining parameters of the model were kept at fixed values, as show in  Table~\ref{tab:sims_params}. For each set of model parameters and exposure times, we run 50 simulations. Each simulation was then fitted using the same model and priors used in our BXA fitting procedure (Sect.~\ref{xfit}). To quantify the reliability of the method we calculated the match percentage for $\log \mnh$ and $\log \LX$, i.e the percentage of simulations were the 90 per cent credible interval estimated using the posterior distribution includes the input value of the parameter for the simulated spectra.

Figure~\ref{fig:reliability} shows our match percentage results for the Hydrogen column density and the intrinsic X-ray luminosity (black circles) for the three $\log \mnh$ values used in the simulations, at different values in the observed counts in the simulated spectra. The overall match percentage (gray triangles) for both parameters is $\gtrsim95$ per cent. The $\log \mnh$ reliability for low or CT absorption is quite high, very close to 100 per cent, even when the counts in the spectrum are below five counts. As we showed in the previous section, when the spectral counts are very low the posterior distribution are poorly constrained compared with the initial prior, so the estimated confidence intervals are very large. In general, for low or CT absorption level the posteriors are not strongly constrained, giving at best an upper or lower limit estimate, respectively. For $\log \mnh=23.5$ the match percentage is lower. Giving the redshift and observed energy range of the simulations, at this absorption level the posterior are better constrained, with narrower confidence intervals. The drop in reliability is most significant between 10 and 50 counts, but even in this case the match percentage is close or above 90 per cent.

A more detailed analysis would be needed for a complete characterization of the reliability of the UXClumpy model using BXA spectral fits (see e.g. \citealt{Saha2022}), including a full exploration of the parameter space and more realistic simulations taking into account the effects of the background emission and different instrumental setups. Nevertheless, the results we presented here show enough evidence to conclude that no significant biases affect our estimations of the Hydrogen column density or the intrinsic X-ray luminosity, even when low-count spectra are considered.

\end{appendix}

\end{document}